# Knowledge-Guided Construction of Machine Learning Potentials for Reactive Behavior in Energetic Materials

*Guanchen Dong[1,2,3,4], Huanrong Liu[1], Lin Wang[1], Yuxuan Chen[1,3], Zhengyu Tu[1,3], Qing-an Li[1,3], Bin Xu[1], Jian Li[1], Shan Zhang[1], Xiaotong Li[1], Linghua Tan[2], Guangcheng Yang[2,4], Jun Kang[3], Rui Su[5*], Pengfei Guan[1,3*]*

1. Advanced Interdisciplinary Science Research (AiR) Center, Ningbo Institute of Materials Technology and Engineering, Chinese Academy of Sciences, Ningbo 315201, China
2. School of Chemical Engineering, Nanjing University of Science and Technology, Nanjing, 210094, China.
3. Beijing Computational Science Research Center, Beijing 100193, China
4. China Academy of Engineering Physics·Institute of Chemical Materials, Mianyang, 621900, China.
5. Institute of Advanced Magnetic Materials, College of Materials & Environmental Engineering, Hangzhou Dianzi University, Hangzhou 310018, China

**Corresponding Author**

* Rui Su (surui@hdu.edu.cn); Pengfei Guan (pguan@nimte.ac.cn).

**ABSTRACT**

Energetic materials involve multiple interactions, making reliable machine-learning potentials (MLPs) difficult to construct without extensive datasets. Here, we enhance the ability of the neural network interatomic potential (NNAP) to describe molecular systems using sphmulti, a refined spherical-harmonic basis set that combines extra radial basis with element-specific chemical weighting in each radial channel. We further integrate hyperactive learning, farthest-point sampling, and iterative sufficiency criteria with knowledge-guided key events, enabling the NNAP to retain essential information for describing the material system. For 3,3'-diamino-4,4'-azoxyfurazan, the resulting NNAP is trained on only 13,680 configurations yet captures key properties, including the isomerization barrier, pyrolysis activation energy, and shock Hugoniot behavior. NNAP-driven molecular dynamics simulations further reveal that pyrolysis proceeds through furazan-ring opening followed by chain reactions, whereas shock loading induces anisotropic hydrogen transfer driven by interlayer structural deformations. This work enables efficient and reliable MLPs for energetic materials.

Energetic materials (EMs) possess exceptionally high energy densities, offering substantial potential for societal benefit but simultaneously posing serious challenges for safety control[1–3]. A fundamental understanding of the mechanisms governing energy release (i.e., explosion) is therefore critical[4]. Typically initiated by external stimuli, explosive processes span multiple scales, and resolving these dynamics is essential for the accelerated development and rational design of advanced EMs[5,6]. Under extreme physical conditions, explosive responses depend sensitively on the stimuli and material type, leading to diverse reported initial reaction pathways for representative EMs (Table S1); the lack of clear prior knowledge therefore substantially prolongs the path from candidate materials to practical applications[7]. For example, 3,3'-diamino-4,4'-azoxyfurazan (DAAF) exhibits the unusual combination of impact insensitivity and shock sensitivity[8,9]. Elucidating this behavior of such insensitive energetic materials requires probing structural and chemical transformations over timescales extending to nanoseconds[10]. Trial-and-error experimentation with EMs is inherently hazardous, while conventional pathway search methods, such as transition state theory, remain computationally expensive for comprehensively exploring reaction networks or identifying dominant pathways in the maze-like dynamics of explosive processes[11–13]. Over the past decades, in silico simulations, particularly molecular dynamics (MD), have become indispensable for investigating the actual dynamical processes in EMs[14]. Quantum-mechanical (QM) methods are one main approach, which are computationally expensive at each time step, making sufficiently long and ergodic simulations of large systems involving rare events prohibitively costly without metadynamics or other enhanced sampling methods[15,16]. The second main approach is based on empirical force fields, which enable simulations of larger and more complex systems[17,18]; however, their accuracy is constrained by the choice of functional form and parameterization, which are highly depend on the material systems and require extensive human intervention[19,20]. Furthermore, the underlying functional forms of empirical force fields struggle to simultaneously capture the non-reactive crystalline features and the reactive dynamics of EMs[21,22].

Capitalizing on recent advances in the application of data science tools to the physical sciences, machine-learning potentials (MLPs) employ physics-agnostic models trained on accurate potential energy surfaces (PES), enabling energy-force predictions with high accuracy. The MLPs in the field of metals and alloys have undergone rapid development[23–27]. However, constructing a reliable MLP for molecular crystals remains challenging, because they involve multiple types of interactions, including covalent bonding, van der Waals interactions, and hydrogen bonding, which together give rise to a complex PES[28–31]. At the same time, high-quality benchmark data remain limited for several key properties, particularly in energetic materials. Nevertheless, the construction of molecular crystal's MLPs can rely on *ab* initio molecular dynamics (AIMD) sampling, supplemented by active learning (AL) strategies (AIMD+AL)[20,32–38]. Since the configurational space for molecular crystals' sampling is weakly constrained a priori, QM labeling can become prohibitively expensive. Moreover, energy and force errors on test sets provide basic metrics of whether an MLP achieves the corresponding QM-level accuracy[39]. However, these metrics cannot by themselves establish whether an MLP can reliably capture the key events relevant to the target material behavior[40]. The reliable MLPs depends on the adequacy of the sampling strategy in encompassing diverse atomic environments[41–44]. Alternatively, it can rely on the use of high-dimensional descriptors (e.g., angular order parameters, message passing neural networks)[31,45–49] to accurately capture multi-scenario dynamic processes, but this reduces efficiency and makes the trade-off between computational cost and accuracy more pronounced.

In this work, we propose a knowledge-guided framework for constructing MLPs of molecular crystals, particularly in EMs. In this framework, we introduce sphmulti, a refined spherical-harmonic representation within our in-house neural network inter-atomic potential (NNAP)[50]. By combining radial engineering of extra radial basis with element-specific chemical weighting in each radial channel, sphmulti provides a more expressive description of molecular systems across distance scales. Moreover,sampling is organized around material key events, thereby ensuring that the resulting MLP has the necessary capability to capture the behaviors relevant to the applications of interest. By integrating hyperactive learning (HAL)[51] with farthest-point sampling (FPS)[52], the sampling strategy efficiently drives exploration beyond local minima and comprehensively captures PES information around the key events. On this basis, we show that DAAF can be comprehensively described with 13,680 configurations, where a single NNAP simultaneously reproduces its structural, mechanical, vibrational, and reactive properties. Furthermore, the NNAP enables the identification of distinct initial reaction pathways under thermal and shock loading. These results establish a reliable framework for investigating the dynamics of EMs under extreme conditions.

## Results

### Improved Machine Learning Potential Basis Set for Molecular Systems

Accurately capturing the dynamics behavior of DAAF under extremely conditions at the atomic scale is pivotal to understanding its unique safety profile. However, progress has long been limited by the lack of reliable potentials. Although existing MLP basis sets have been applied to molecular crystals, achieving high accuracy remains challenging because these systems involve a complex interplay of covalent and noncovalent interactions, often leading to energy mean absolute error (MAE_E) above 10 meV.[33,34,36,38] Optimizing the basis set for such heterogeneous interactions is therefore a key prerequisite for constructing reliable MLPs for molecular systems. DAAF is a layered molecular crystal with an interlayer spacing of approximately 3-4 Å.[53] A conventional radial cutoff of about 5 Å for local atomic-environment basis set can in principle include intermolecular interactions. For MLPs, however, the more critical challenge is that these intermolecular interactions are much weaker than intramolecular covalent bonds. Simply increasing the number of neurons or network layers, or adding a limited number of closely packed configurations to amplify intermolecular interactions, is still insufficient for enabling an MLP to accurately capture intramolecular and intermolecular interactions simultaneously.

To address this issue, introducing radial descriptors with an extended cutoff radius represents the most straightforward strategy.[54] A long-cutoff radial function can incorporate more distant atomic environments in an approximately two-body manner and, when combined with the nonlinear fitting capability of neural networks, enhance the ability of the basis set to represent medium-range interactions. However, an indiscriminate extension of radial cutoffs would substantially increase the cost of descriptor calculation and may introduce redundant features into the neural-network fitting process, thereby deteriorating both accuracy and generalizability of MLP.

To this end, we introduce an additional independent long-cutoff radial basis, referred to as "block2", to specifically describe medium-range dispersion interactions and hydrogen bonding in molecular systems. The original sph is retained as "block1", and the combination of "block1" and "block2" forms the proposed sphmulti basis set. Specifically, "block2" is developed by extending the formulation of the original sph with a larger radial cutoff, thereby complementing the sphmulti with enhanced sensitivity to medium-range interactions. In addition, a covalent-radii-based compression function is introduced in the form of a Fermi function to smoothly modulate the contributions from distant neighbors within the long-cutoff region. This compression function controls the effective interaction range and determines the actual contribution of distant neighbors to the descriptor.

Furthermore, in "block2", the original weighting term is reformulated as an element-dependent weight coupled to the radial channels. The optimized weighting term is constructed using orthogonal sinusoidal features, with the weighting parameters from "block1" embedded as prior information. This design enables an adaptive representation of element-dependent weights, meaning that the weighting term in the independent radial basis does not require data-driven training. By further exploiting radial-channel information, the proposed weighting term improves the accuracy of medium-range interaction descriptions, while efficiently handling multicomponent chemical environments within the long-cutoff region and reducing unnecessary computational overhead.

The introduction of "block2" is expected to incur only a minor additional computational cost, while significantly enhancing the ability of machine-learning potentials to learn how medium-range interactions affect intermolecular mechanical deformation. The resulting sphmulti basis set enables interactions of different strengths between different element pairs to be effectively represented within a unified MLP. It also provides improved sensitivity to both short-range and medium-range structural features, making it particularly suitable for molecular and molecular-crystal systems where intramolecular and intermolecular interactions coexist. Fig. 1 provides a schematic illustration of the basis set improvements, with details given in the "NNAP framework and training details" section.

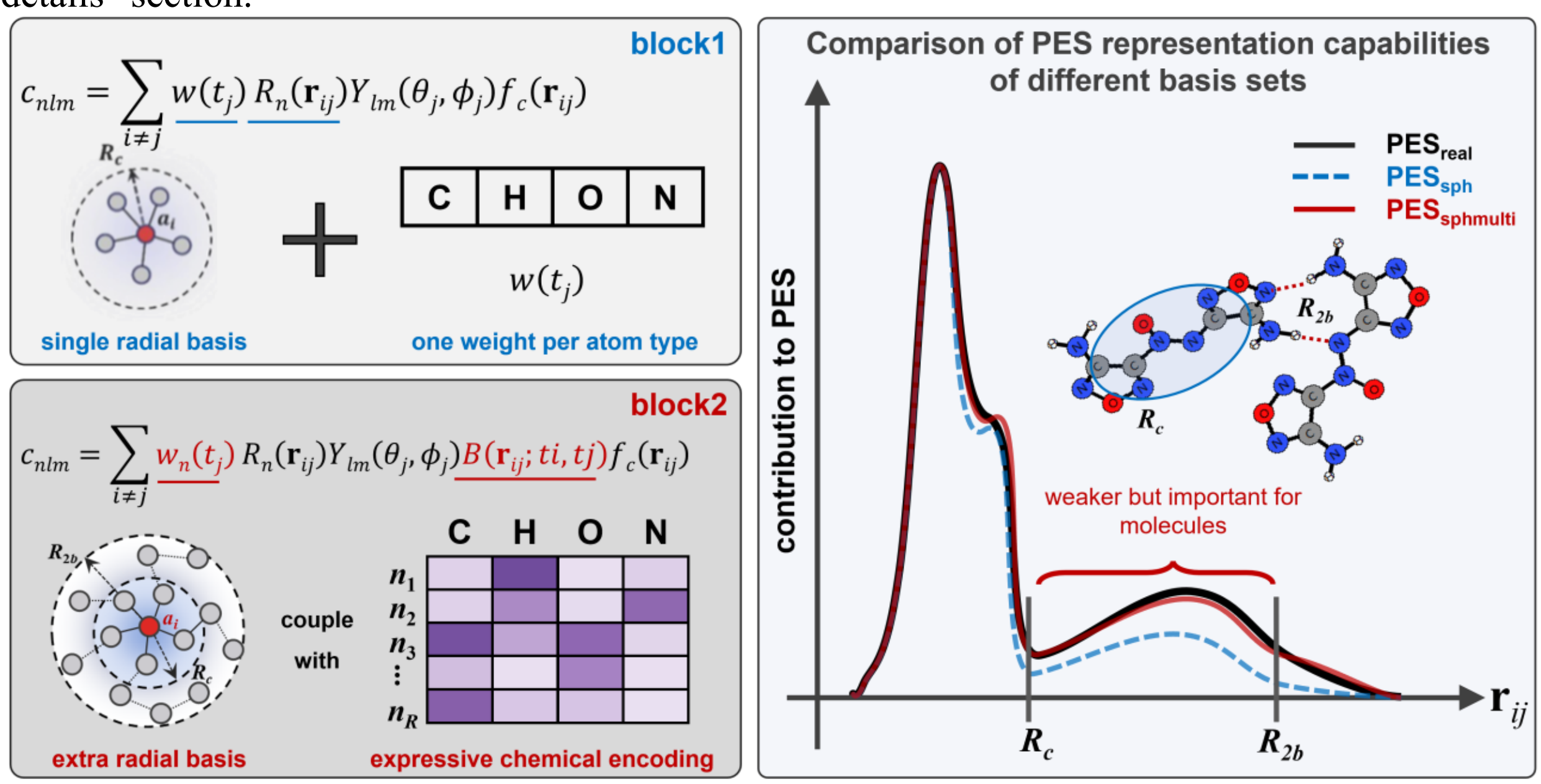

**Fig. 1. Basis set improvement of the NNAP for molecular systems.** The original sph, as "block1", uses a single radial basis together with one weight parameter for each atom type, which mainly captures strong short-range interactions but provides limited resolution for more complex molecular environments. In the improved sphmulti scheme, "multi" refers to the additional extra radial basis "block2" appended to the "block1". An extra radial basis is introduced to better resolve intermolecular interactions, especially in the medium-range regions relevant to molecular packing and non-bonded contacts. Meanwhile, the radial-channel-resolved chemical weighting term assigns distinct chemical weights to different radial channels, enabling expressive chemical encoding. When coupled with the covalent-radius-based compression function, this scheme enhances the ability of the extra radial basis to distinguish interactions between different element pairs. These modifications provide a more expressive representation of local

atomic environments and lead to improved accuracy in molecular energy and force predictions, particularly by improving the description of weaker but important medium-range interactions for molecules.

**Knowledge-guided Machine Learning Potential Framework**

With a suitable molecular-system basis set of MLP, the configurational space of molecular crystals remains vast. We therefore introduce a knowledge guided HAL strategy (Fig. S1). Rather than relentlessly pursuing broad dataset, we decompose the application space of DAAF into the key physical and chemical events that matter in practice and build a target-driven dataset around these events. The goal is to steer model optimization toward faithfully reproducing the dynamics of the key events of interest.

Specifically, we organize sampling into four progressive stages (Fig. S2). This design proceeds from the molecule to the crystal, then to pyrolysis, and finally to shock stage, so as to progressively establish the description of the DAAF system and improve sampling efficiency in the later stages. At each stage, the sampling design is paired with an explicit criterion for iterative sufficiency criteria based on the targeted key properties. HAL is coupled to each stage to enrich the dataset with high-uncertainty configurations, while FPS is further used to reduce the cost of density functional theory (DFT) labeling by preserving representative structural diversity among the selected configurations. Through this progressive and knowledge-guided procedure, the resulting NNAP acquires the necessary capability to capture the complex reactive behavior of DAAF.

DAAF is a layered EM molecular crystal, composed of two furazan rings linked by an azoxy group. NNAP was first endowed with the basic capability to capture key molecular property, and the sampling was then extended to the crystal stage, which improves sampling efficiency and yields the NNAP with better generalizability on the DAAF crystal. For energetic materials, both thermal and shock responses are essential for detonation design. In the pyrolysis stage, the activation energy ($E_a$), which is closely related to pyrolysis kinetics, was used as the targeted key property. In the shock stage, where the response is governed by the thermal excitation and mechanical deformation, the key event was defined as the macroscopic behavior in the elastic regime, i.e., the shock Hugoniot relation. This choice is justified because the thermal-related $E_a$ had already been sufficiently constrained in the preceding pyrolysis stage. The detailed sampling protocols and reproducible sampling workflow are shown in Fig. 2a.

A stage is considered converged when the knowledge-guided metric is met and remains stable upon adding more data. As summarized in Fig. 2b-e, for the molecule stage, the deviation of cis–trans isomerization barrier between the NNAP and DFT is less than or equal to the MAE_E of test set. For the crystal stage, the NNAP-driven biased MD remains stable, and its stability metric no longer exhibits noticeable changes with additional training data. For the pyrolysis stage, the $E_a$ predicted by NNAP falls within the experimentally reported range; the final value obtained by NNAP is 133 $kJ \cdot mol^{-1}$, in good agreement with 135 $kJ \cdot mol^{-1}$ reported by Gao et al[55]. For the shock stage, the Hugoniot predicted by NNAP reproduces the experimental trend. Specifically, the LANL measurements[56] on DAAF containing 5% Kel-F give the empirical relation $U_s = 2.51+1.79\ U_p$ (where $U_s$ is shock velocity and $U_p$ is particle velocity), while our NNAP-based Multi-Scale Shock Technique (MSST)[57] of perfect crystalline DAAF yield a fitted Hugoniot of $U_s = 3.02+1.62\ U_p$. Once the convergence metric for a given stage is satisfied, the framework exits the current stage and proceeds to the next. A quantitative summary of the NNAP performance for the key events considered in this work is provided in Table S3. Beyond targeted evaluations on key events of DAAF, we carried out a comprehensive benchmark of NNAP on generic properties of crystalline DAAF and compared it against available force fields and experiments. Whether in terms of accuracy or efficiency, NNAP performs better than DAAF-specific ReaxFF[58], EMFF2025 (trained on CHON energetic-material dataset based on DeePMD)[38], and fine-tuned NEP89[59] (details see Method) under the same test conditions.

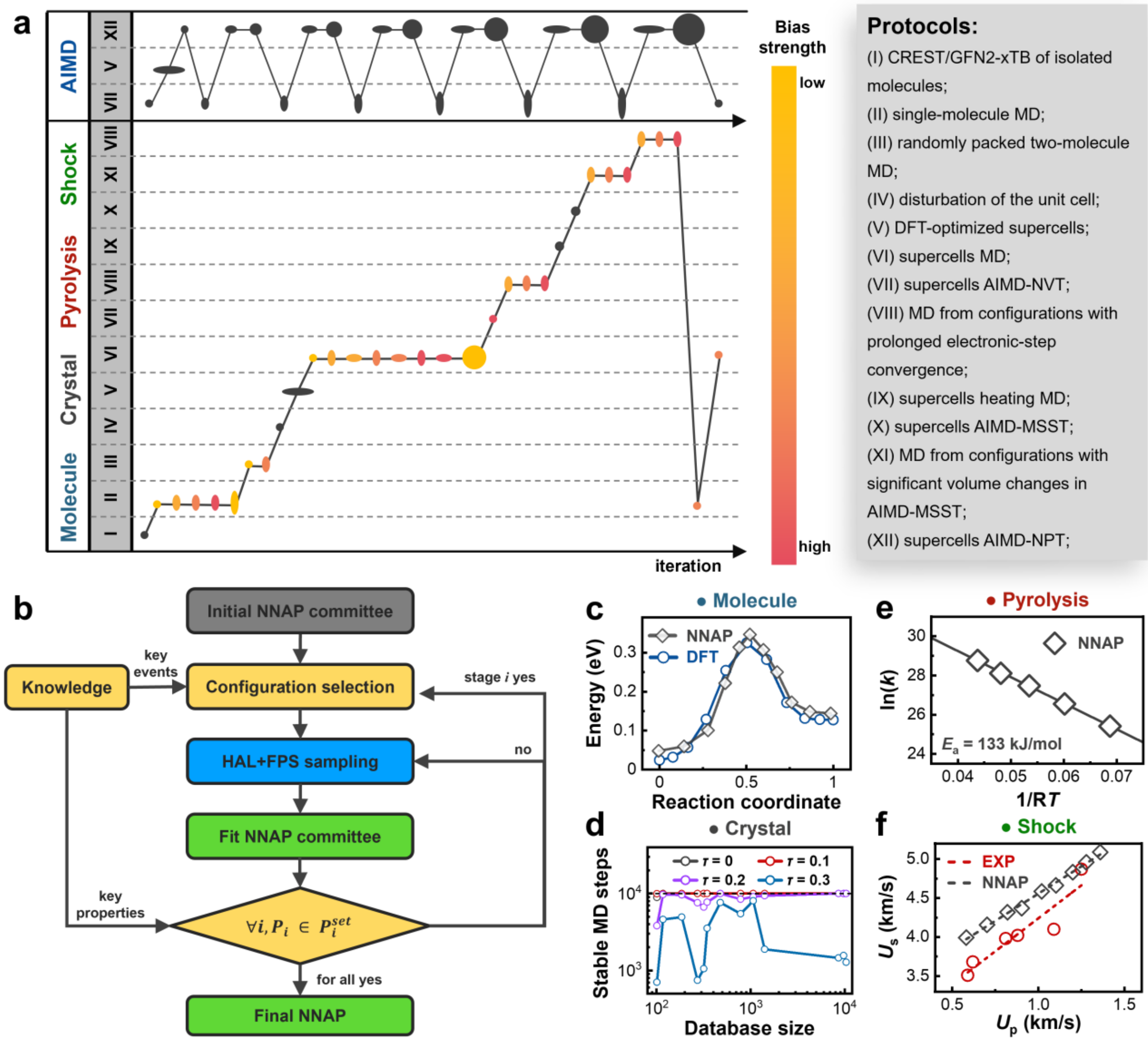

**Fig. 2. Knowledge-guided and reproducible workflow for reliable MLP construction.** (a) MLPs construction through knowledge-guided stage-specific (NNAP) and AIMD+AL sampling strategy (NNAP_AIMD). The NNAP is constructed using a progressive knowledge-guided framework. First, key events are defined based on knowledge, and the relevant conformations are selected. At the same time, the criteria $P_i^{set}$ for determining when the iteration for that stage is complete are established based on the key properties corresponding to those key events. Once all the criteria have been met, the final NNAP is obtained. Specially, the NNAP is progressively constructed through four stages with the following sampling protocols. The sampled configurations are then deduplicated and screened according to the filtering criteria described in the section "Knowledge-guided initial configuration selection and iterative criteria setting" within SI, and important configurations are retained for MLP construction. Here, the variable in optimization is pressure; in MD are pressure (vertical length), temperature (horizontal length), and bias strength (color weight), the heating MD excludes. In addition, supercells refer to 2×1×1 supercells of in the crystal stage or AIMD sampling strategy, and 2×2×1 supercells in the pyrolysis and shock stages. Different velocities and loading directions are adopted in AIMD-MSST. Configurations of (VIII) in the pyrolysis stage from AIMD-NVT and in the shock stage from AIMD-MSST. (b) Schematic of knowledge-guided construction of machine learning potentials. (c) Energy barrier for cis-trans isomerization of isolated DAAF. (d) Stability metric for NNAP-driven biased MD versus dataset size. (e) Pyrolysis activation energy of crystalline DAAF predicted by NNAP. (f) Hugoniot of crystalline DAAF predicted by NNAP.

To demonstrate the advantages of the improved basis set combined with the knowledge-guided sampling strategy, Fig. 3 compares the test set accuracy of MLPs constructed with different basis sets and sampling strategies. Tested on the same dataset, the basis set-improved $NNAP_{sphm}$ reduced the MAEs of energy and force by 27.4% and 24.4%, respectively, while lowering the root mean square errors (RMSEs) by 23.2% and 21.3% relative to original basis set sph ($NNAP_{sph}$), demonstrating the effectiveness of the basis set improvement for molecular systems. We also tested $NNAP_{sphl}$, which adds an extra radial basis to "block1" while retaining the original weighting term of sph. Fig. 3 shows that $NNAP_{sphl}$ merely extending the radial cutoff remains less effective than $NNAP_{sphm}$. As shown in Fig. S8, although $NNAP_{sph}$ is 2.01 times more computationally efficient than $NNAP_{sphm}$, its accuracy is considerably inferior. Moreover, compared with $NNAP_{sphl}$, which only extends the radial cutoff, $NNAP_{sphm}$ exhibits superior performance in both accuracy and computational efficiency, indicating that the proposed improvement is highly effective and incurs only an acceptable additional computational cost. Moreover, although the NNAP_AIMD (see the "Machine learning potential (MLP) constructed using the AIMD+AL strategy" in "Suppelementary Information (SI)" for details.) with an exceptionally low force errors of MAE (MAE_F) of 0.0476 $eV \cdot Å^{-1}$ on test sets (the partitioning of the training/validation/test sets is described in the "NNAP framework and training details"), the resulting NNAP_AIMD may still be inadequate for describing the dynamical behavior of the DAAF molecular crystal. (Fig. 3e, f and Table S2). The main reason is that the sampled configurations are largely confined to local minima, leading to insufficient coverage of the physically relevant configurational space[60]. Consequently, test sets energy/force errors alone are inadequate for assessing whether an MLP can describe the fundamental properties of the material system[40]. This framework establishes the necessary foundation for the NNAP to reproduce key events, while drastically reducing the DFT cost relative to AIMD+AL strategies. The conventional AIMD+AL sampling spreads uniformly in the principal component space of descriptors, which poorly covers the critical regions associated with shock and pyrolysis (Fig. S3). This contrast underscores that low test sets error is insufficient; a knowledge-guided construction of MLPs is essential for capturing the intrinsic properties of the material system.

Furthermore, in the section "Robustness Benchmarks of NNAP" within the SI, Figs. S4-7 are supplemented. Comprehensive benchmarks are systematically conducted from the perspectives of energy and force errors on the test set, cell parameters, mechanical properties, and thermodynamic information. These highlight the rationality of the knowledge-guided MLP construction and the high efficiency of the NNAP basis set.

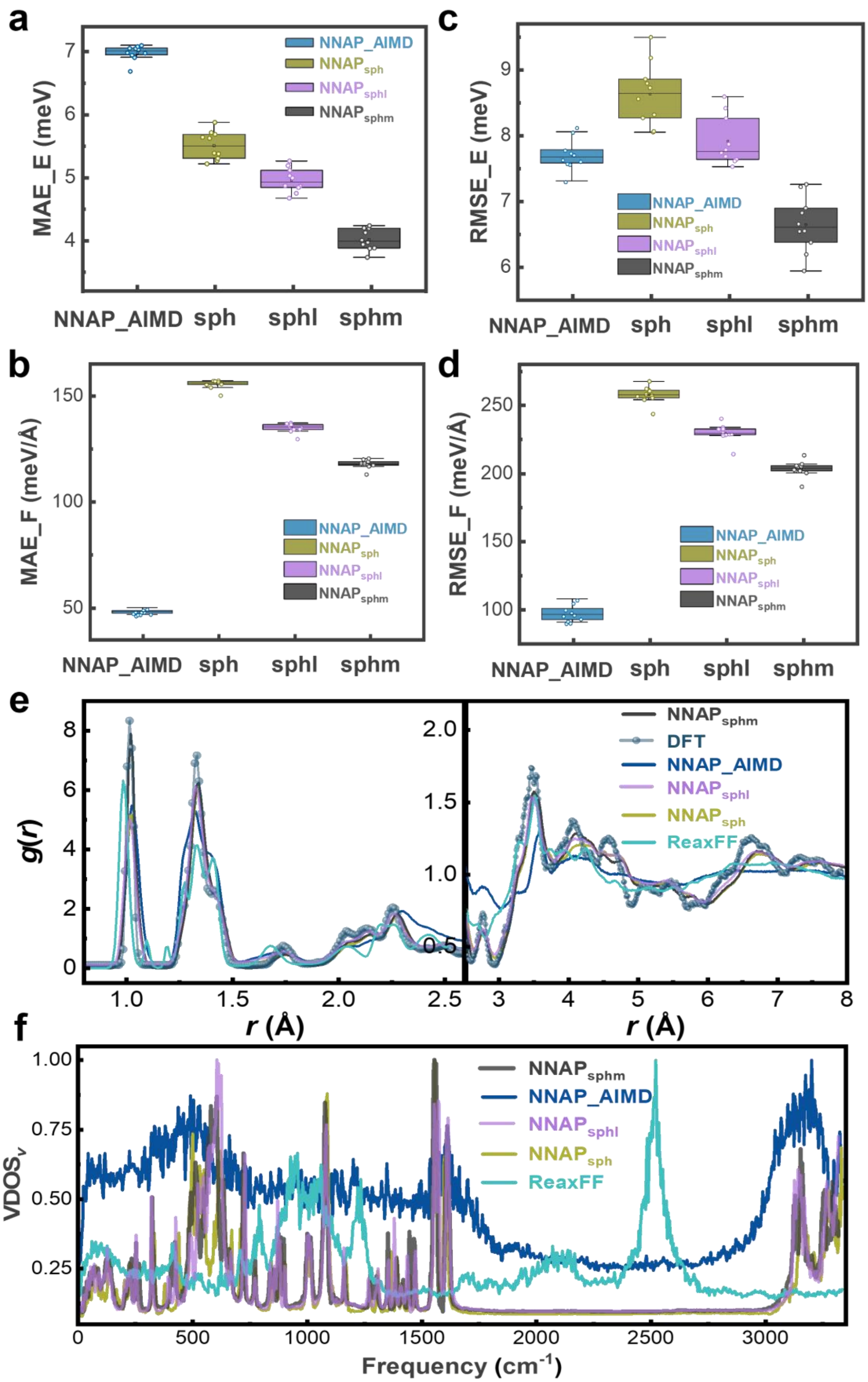
a
MAE_E (meV)
NNAP_AIMD
NNAP_sph
NNAP_sphl
NNAP_sphm
NNAP_AIMD sph sphl sphm
b
MAE_F (meV/Å)
c
RMSE_E (meV)
d
RMSE_F (meV/Å)
e
g(r)
DFT
ReaxFF
r (Å)
f
VDOS_v
Frequency (cm^-1)

**Fig. 3. Benchmark of NNAP constructed with improved basis set and knowledge-guided sampling.** Energy/force error ranges on the test sets obtained by 10-fold cross-validation for potentials constructed using different sampling strategies or basis sets. The dashed lines indicate the average errors. (a, b) The MAE_E/F is the energy/force errors of mean absolute error, and (c, d) the RMSE is the root mean square error. (e) RDFs of DAAF for $r \leq 2.5$ Å and $2.5 \leq r \leq 8.0$ Å, calculated from structures in the 300 K NPT ensemble using different potentials. (f) VDOSs of DAAF calculated from atomic velocities obtained from 300 K NVE simulations using different potentials.

## Initial Pyrolysis Reaction Pathways

The microscopic details of pyrolysis in EMs are notoriously difficult to access experimentally Leveraging our high-fidelity NNAP (Hereafter, we refer to $NNAP_{sphm}$ simply as NNAP.), we perform large-scale MD simulations and statistically resolve the dominant initial reaction pathways of DAAF under pyrolysis.

To track the complex reaction network of DAAF decomposition, we developed a bonding- and topology-based species identification scheme that explicitly distinguishes DAAF possible furazan ring opening (RO) and hydrogen transfer (HTr)[61] events; detailed criteria are provided in the "Species identification" section. As shown in Fig. S9, we find pyrolysis that is predominantly initiated by the RO, accompanied by the cleavage of the C–N bonds connecting the furazan ring and the azoxy group. The early production is dominated by NO, which subsequently evolves toward $N_2$ (consistent with the experimental results[62] obtained using excited electronic state decomposition, which suggests that NO is the initial small molecule production of DAAF. And it is only produced in the early stages and quickly converted into other species during pyrolysis). $NH_3$ is also notable small-molecule product generated during pyrolysis process. The overall species evolution exhibits weak temperature dependence (Fig. S10) and is consistent with 3000 K AIMD trajectories (Fig. S9a, inset). Notably, although our training set did not explicitly include isolated small-molecule products or pre-seeded RO configurations, NNAP successfully reproduces these response processes, highlighting the strong extrapolative capacity imparted by our knowledge-guided sampling strategy.

Guided by these MD statistics, we identify two main initial reaction pathways (Fig. 4a,b): (i) RO: This is the most dominant initial reaction mechanism under thermal stimuli. In greater detail, we find that RO is driven primarily by the cleavage of the N–O bond within the furazan ring, distant from the azoxy group. This is coupled with azoxy cleavage, which occurs mainly through asymmetric azoxy cleavage (AAC) and secondarily through symmetric azoxy cleavage (SAC). Once RO occurs, a chain of events is rapidly triggered, including NO formation. (ii) HTr: This pathway primarily involves the intermolecular transfer of H atoms from amino groups to furazan-ring N atoms in neighboring molecules, with only a minor fraction transferring to the azoxy oxygen.

To thermodynamically validate the initial pyrolysis reaction pathways inferred from MD statistics, we perform Nudged Elastic Band (NEB) calculations at both NNAP and DFT levels, comparing the occurrence relative ease of these representative reaction pathways, RO and HTr (Fig. 4c). For RO, we design two pathways: AAC and SAC. For HTr, we design three possible transfer pathways based on the amino group's closest approach. Specifically, for the amino group adjacent to the azoxy O atom, we consider inter-layer (HTr1) and intra-layer (HTr2) transfer pathways, while for the amino group remote from the azoxy O atom, we consider intra-layer transfer (HTr3). The initial configurations for NEB were selected from representative structures extracted from the pyrolysis trajectory. The order of reaction difficulty revealed by NNAP aligns with that predicted by DFT. AAC emerges as the most thermodynamically favorable and hence the highest frequency initial chemical reaction of RO under thermal stimuli. By contrast, SAC proceeds through a higher-energy intermediate and tends to undergo C–N[O] bond cleavage, which can readily evolve into the same ring-opened state as AAC (RO-AAC). HTr processes all exhibit relatively high energy barriers, with HTr3 being the most likely to occur. The above ranking of reaction tendency explains the species-evolution trends in Fig. S9 and provides thermodynamic support for the initial pyrolysis reaction pathways identified by NNAP.

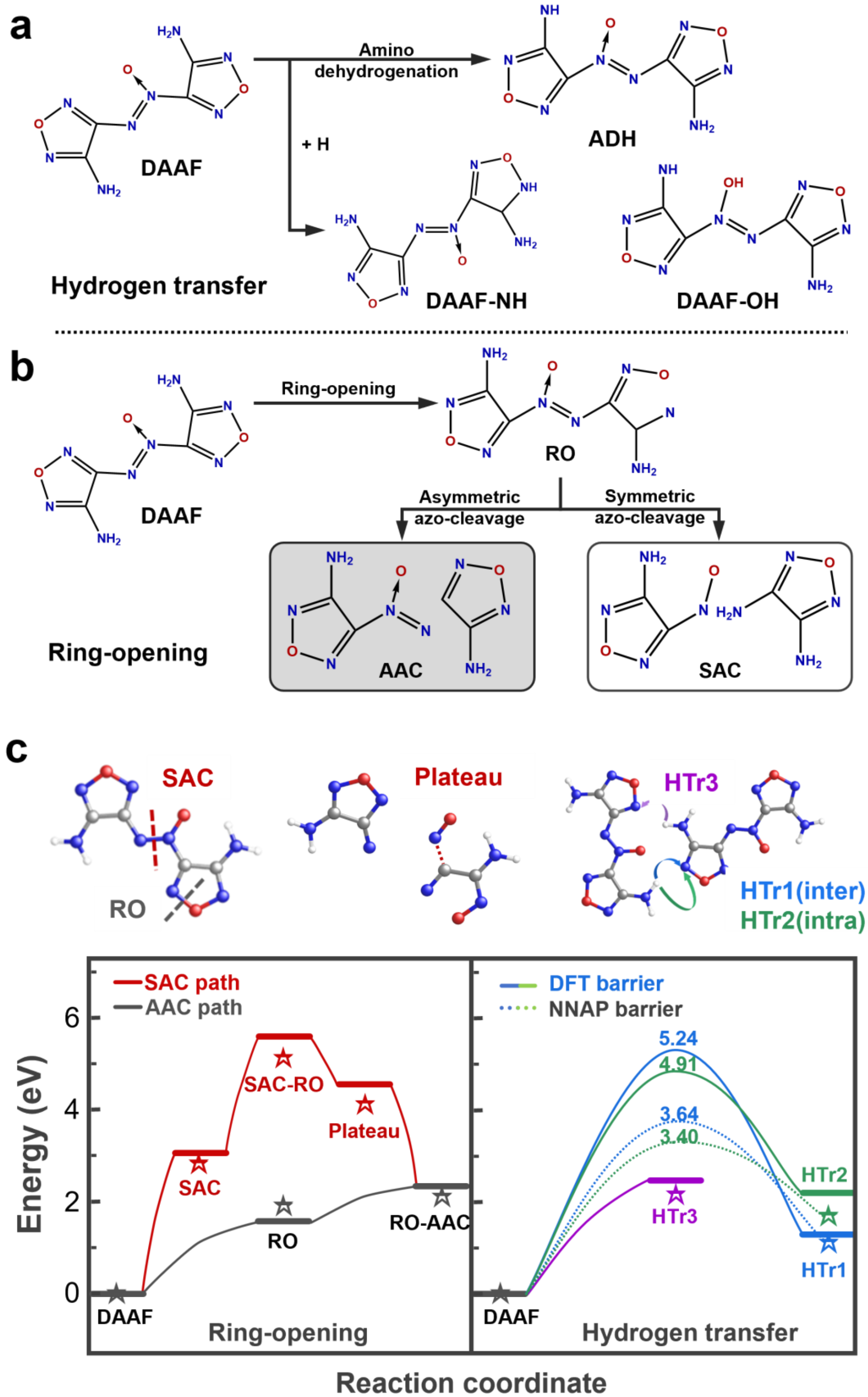
a
Amino
dehydrogenation
+ H
DAAF
ADH
DAAF-NH
DAAF-OH
Hydrogen transfer
b
Ring-opening
RO
Asymmetric
azo-cleavage
Symmetric
azo-cleavage
AAC
SAC
Ring-opening
c
SAC
RO
Plateau
HTr3
HTr1(inter)
HTr2(intra)
SAC path
AAC path
SAC-RO
Plateau
SAC
RO
RO-AAC
DAAF
Ring-opening
DFT barrier
NNAP barrier
5.24
4.91
3.64
3.40
HTr2
HTr3
HTr1
DAAF
Hydrogen transfer
Energy (eV)
Reaction coordinate

**Fig. 4. Initial pyrolysis reaction pathways of DAAF and thermodynamic support from NEB.** (a) HTr pathway: initial dehydrogenation occurs from amino groups, with the detached H atoms transferring primarily to the N atom of the furazan ring, and to a lesser extent to the O atom of the azoxy group. (b) RO pathway: cleavage of the N–O bond in the furazan ring distant from the O atom of the azoxy group initiates ring opening, followed predominantly by AAC and secondarily by SAC. (c) Schematic of RO- and HTr-related reaction pathway explored by NEB. Energy variations along these pathways are shown for NNAP (stars) and DFT (plane).

## Anisotropic Response under Shock Loading

Direct experimental or AIMD resolution of shock-initiated chemical events in DAAF remains elusive, because of the extreme conditions and the spatiotemporal scales of detonation. The behaviour of NNAP during the pyrolysis simulation provides a kinetic benchmark for subsequent chemical reactions under shock loading. Furthermore, NNAP reproduces the experimental trends observed in the Hugoniot of DAAF and further recovers the DFT-level energy evolution during shock loading of small-scale configurations (Fig. S11). Together, these benchmarks demonstrate that NNAP has the necessary capability to capture the response behavior of DAAF under shock loading. We therefore employ NNAP to investigate shock loading in DAAF along characteristic directions relative to the molecular layers. Shocks are launched along the $x$, $y$, $z$ axes of our orthogonalized simulation cell (Fig. S12a). In brief, $z$ is the layer normal, while $x$ and $y$ lie within the molecular layers. The evolutions of temperature, pressure, and density are closely matched for $x$/$y$/$z$ loading (Fig. S12b-d), ensuring that any chemical differences reflect orientation-dependent structural response rather than thermodynamic disparities.

The chemical reaction response of DAAF under shock loading is fundamentally different from pyrolysis. Although both involve RO and HTr, the HTr becomes significant during shock. Amino groups rapidly generate H• radicals, which drive cluster formation, and conversion of fragments to more stable small molecules. By contrast, RO serves mainly as a structural accommodation that stabilizes clusters and enables growth to larger clusters, as evidenced by the paucity of NO (Fig. 5a and S13).

Moreover, the initial response is also anisotropic, especially in the deformations of the molecular layers. The statistical analysis of structurally distinct key intermediates in Fig. 5b enables the extraction of representative chemical reaction pathways from the shock trajectories. Shock loading along the $x$ direction, pronounced furazan-ring torsion and C–$NH_2$ bending produce local folding near the azoxy and amino groups, shortening the transfer distance for HTr to ring N. The high-frequency chemical reactions in this case are dehydrogenation and H transfer to ring N, followed by rapid polymerization (Fig. 5c). Shock loading along the $y$ direction causes neighboring molecular sheets to rotate in opposite senses about the center, forming an X-like pattern in the $x$–$z$ plane while still maintaining the integrity of the molecular layers. This rearrangement resembles an alternation displacement, which brings amino groups from neighboring molecules into closer proximity and facilitates inter-amino H transfer (Fig. 5d). In contrast, under $z$ loading, a dislocation-like deformation combined with interlayer slip and intra-furazan bending makes RO most frequent, and thus RO-driven copolymerization emerges as the dominant chemical reaction pathway in this case (Fig. 5e).

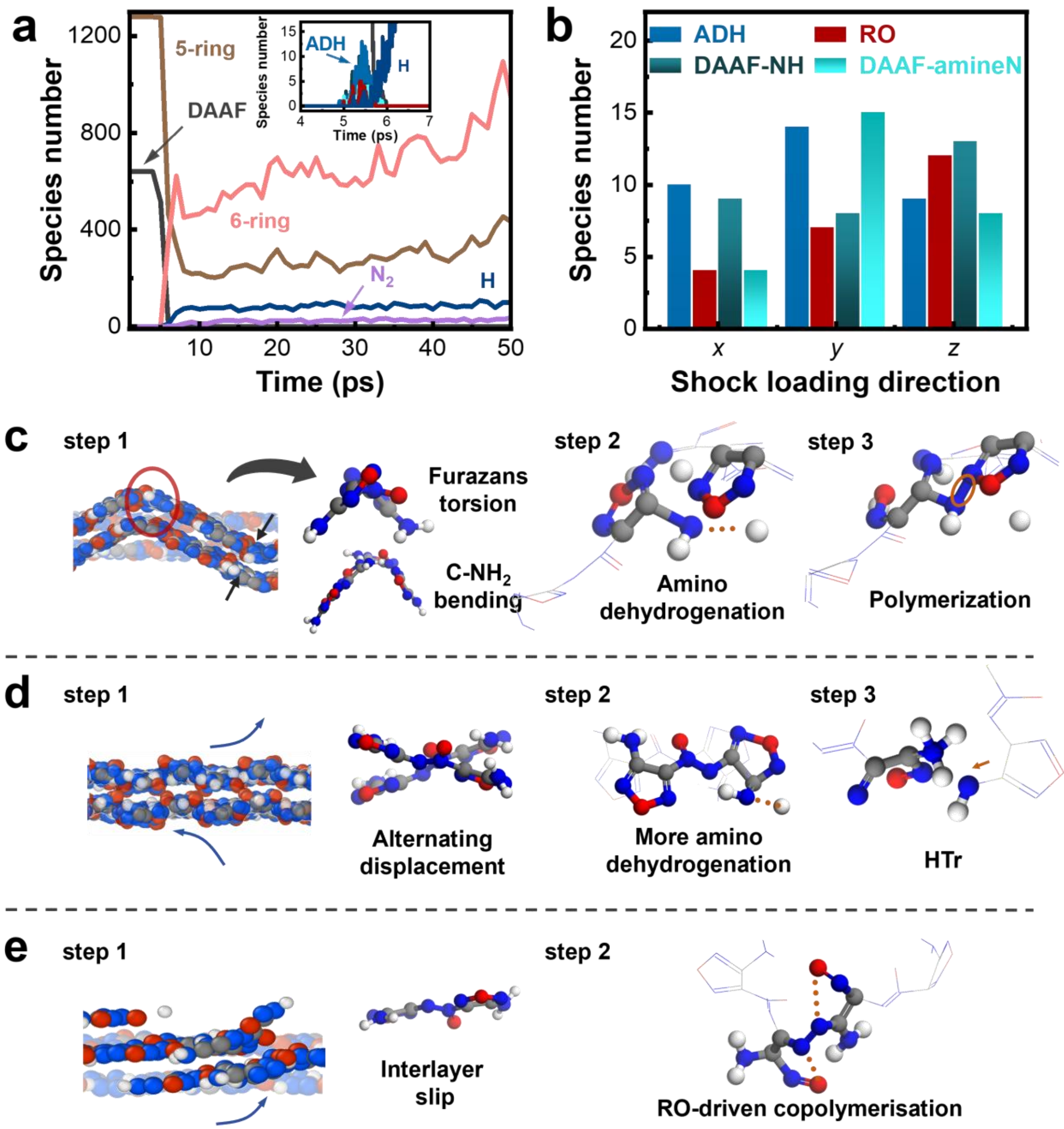

**Fig. 5. Shock-induced initial reaction pathways and anisotropy in DAAF.** (a) Time evolution of the major species under shock loading along the *x* direction; inset shows detailed counts of key intermediate species. (b) Peak counts of key intermediates under shock loading along different directions. DAAF-amineN denote species with an additional H atom attached to an amino N atom. (c-e) Structural deformations and statistically extracted dominant initial reaction pathway from trajectory under shock loading along different directions.

## Discussion

NNAP, trained within a knowledge-guided HAL sampling framework, achieves the balance between the accurate description of key events and computational efficiency. Rather than focusing solely on minimizing test-set errors, the workflow anchors sampling to toward diverse configurations surrounding key events, ensuring that NNAP captures not only equilibrium configurations but also rare, decisive configurations. On this basis, NNAP resolves—for the first

time—the initial response behaviors and reaction pathways of DAAF under thermal and shock loading within current DFT accuracy.

Compared with widely used empirical force fields and existing MLPs, NNAP shows superior reliability in three key aspects. First, the improved sphmulti basis set enhances the ability of the NNAP to describe molecular systems through radial engineering with radial channel-resolved chemical weighting term, while introducing only an acceptable computational overhead. Second, knowledge-guided sampling guarantees that structural rearrangements and bond-breaking under external stimuli are explicitly represented in the training set, avoiding spurious predictions common to purely data-driven models. Third, the combination of HAL and FPS reduces the amount of data needed to capture key events, thereby decreasing the computational cost of DFT labeling. Together, these features enable NNAP to reproduce complex events with high fidelity and strong transferability.

Statistical analyses from NNAP-MD trajectories further reveal distinct dominant reaction pathways of DAAF under thermal and shock loading. Under pyrolysis, RO dominates the initial pathway and rapidly triggers chain reactions that yield small molecules such as NO and $N_2$. In contrast, under shock loading, HTr emerges as the dominant initial pathway. This shift is driven by pronounced anisotropic structural deformations, which substantially reduce the configurational distances required for different HTr pathways. These findings deliver the first DFT-level resolution of DAAF responses under both shock and pyrolysis, offering a high-fidelity atomic-scale toolset for understanding reactive behavior in EMs.

The framework presented here can be readily extended to other molecular crystals. The major open challenge lies in tailoring sampling strategies to ensure adequate coverage of knowledge-guided key events in diverse materials. Integrating this framework with AI agents holds great potential for automating the training of high-fidelity MLPs, and this will be the focus of future work.

## Methods

### Sampling and filtering strategy

During the NNAP iteration process, we train a 10-member committee of NNAP models using 10-fold cross-validation. The model committee provides an estimate of prediction uncertainty $U$ for each configuration:

$$u_i = \frac{\left\langle \left\| \mathbf{F}_i^{(k)} - \bar{\mathbf{F}}_i \right\| \right\rangle_k}{\left\| \bar{\mathbf{F}}_i \right\| + \varepsilon} \quad (1)$$

$$U = \max_i u_i \quad (2)$$

where $\mathbf{F}_i^{(k)}$ denotes the force on atom $i$ predicted by model $k$, $\bar{\mathbf{F}}_i$ is the committee-averaged force, and $\varepsilon$ is a small regularization constant to prevent division by zero. The physically important extrapolative events are typically local, arising first at a few atoms involved in bond distortion or reaction initiation, whereas averaged atomic deviations or configuration-level force norms can dilute such localized signals.

Biased MD are driven by the committee-averaged forces, with the biased potential energy $E_{\mathrm{biased}}$ defined as

$$E_{\mathrm{biased}} = \bar{E} - \tau \sqrt{\frac{1}{K} \sum_k \left( E^{(k)} - \bar{E} \right)^2} \quad (3)$$

where $E^{(k)}$ denotes the energy predicted by model $k$, $\bar{E}$ is the committee-averaged energy, $\tau$ is bias strength parameter, and $K$ is number of models in the ensemble. Compared with ordinary unbiased MD, where the trajectory evolves tend to remain within already accessible local basins, the bias term provides an additional driving force that gently steers the system toward less-explored regions of configuration space. In practice, the bias does not aim to override the dynamics, but to enhance the probability of visiting configurations where the NNAP's uncertainty is high.

The configurations generated in biased MD are then filtered according to:

(i) the $U$ lies within the setting [0.4, 4.0]; (ii) the z-score ≥ 95% ($\frac{U-\langle U\rangle}{\sigma_U}$ the mean value $\langle U\rangle$ and standard deviation $\sigma_U$ are computed over all previously sampled configurations in the dataset), providing a dynamic statistical baseline for the z-score normalization; (iii) accepted configurations are temporally spaced by a minimum interval to eliminate redundancy; (iv) at each iteration, candidate configurations are ranked in ascending order according to their maximum atomic forces, and the lowest 90% are retained for further screening, and. (v) the retained configurations are further deduplicated by FPS, with the number of selected configurations capped at 200 in each iteration for DFT labeling. After the above filtering steps, the selected configurations are used to refine the NNAP committee.

This knowledge-guided HAL sampling systematically defines a rational and clear direction for the sampling design and iterative sufficiency criteria, thereby avoiding aimless sampling during the MLP construction.

### DFT calculation

All energies and atomic forces used for training were obtained using atomic simulation environment (ASE)[63] and Vienna ab-initio simulation package (VASP)[64]. Before DFT calculations, initial magnetic moments were assigned to each configuration using crystal hamiltonian graph neural network (CHGNet)[65] to facilitate charge and spin convergence. The Perdew–Burke–Ernzerhof (PBE)[66] exchange–correlation functional was employed, combined with Tkatchenko–Scheffler (TS)[67] dispersion correction to account for van der Waals interactions.
For DAAF crystal calculations, convergence tests determined that an energy cutoff of 550 eV was sufficient for the plane-wave basis set. The k-point spacing was set to 0.2 $Å^{-1}$ for crystalline systems, while for isolated systems placed in a 10 Å vacuum layer, a k-point spacing of 50 $Å^{-1}$ was used.

**NNAP framework and training details**

The NNAP framework was built based on our previous work[50] which implements the Spherical Chebyshev basis to build the rotational invariant descriptor through the power spectrum of spherical expansions of neighbor atom positions. Here, we briefly introduce the main formula of the basis. The power spectrum of atom density functions is defined by:

$$P_{nl} = \frac{4\pi}{2l+1} \sum_{|m| \le l} c_{nlm}^{*} c_{nlm} \tag{4}$$

where the atom density function is defined as:

$$c_{nlm} = \sum_{i \ne j} w_n\left(t_j\right) R_n\left(r_{ij}\right) Y_{lm}\left(\theta_j, \phi_j\right) B\left(r_{ij}; t_i, t_j\right) f_c\left(r_{ij}\right) \tag{5}$$

The radial basis and cutoff function are defined in our previous work. Additionally, the $w_n(t_j)$ defined the radial channel-resolved chemical weighting term, which is defined as:

$$w_n^k\left(t_j\right) = \begin{cases} Z_j, & k=1 \\ \sqrt{\dfrac{2}{\max\left(N_r, N_t\right)+1}} \sin\left(\dfrac{(n+1)\pi t_j}{\max\left(N_r, N_t\right)+1}\right), & k=2 \end{cases} \tag{6}$$

where $N_r = n_{max} + 1$ and $N_t$ is the number of atom types. Different $k$ value correspond to independent part of power spectrum, where the final spectrum $P = P_1 \oplus P_2$. The covalent-radii-based compression function $B(r_{ij}; t_i, t_j)$ is defined as:

$$B\left(r_{ij}; t_i, t_j\right) = \frac{1}{e^{\left(r_{ij} - r_0\right)/\sigma} + 1} \tag{7}$$

where $r_0 = b_0(c_i + c_j)$. Here, $c_i$ stands for covalent radii of atom type $i$ , $b_0$ and $\sigma$ are adjustable parameters to finetune the shape of function.
In this work, we further introduce an independent radial basis with larger cutoff $r_{cut}^{(2b)}$ to capture the medium-range information. While introducing only marginal computational overhead, such a hybrid representation enhances the sensitivity to both short-range covalent geometry and long-range structural ordering, which is particularly beneficial for molecular and molecular-crystal systems where inter- and intra-molecular interactions coexist.
The descriptor of this work was constructed with $n_{\max} = 6$, $l_{\max} = 5$ and a radial cutoff of 5.0 Å for the main part of descriptor. An extra radial basis was included with $n_{\max}^{(2b)} = 6$ and $r_{cut}^{(2b)} = 7.0$ Å. The bond parameters were selected as $b_0 = 0.8$ and $\sigma = 1.0$. We construct an atomic neural network for each atom type, which consists of four hidden layers with 40 units per layer. The SiLU (Sigmoid Linear Unit)[68] activation function was used. The atom neural network was trained by the AdamW algorithm[69] with an initial learning rate of 0.002 and an exponential decay factor of 0.95 applied every 500 steps. The training was performed for 4000 epochs with a batch size of 360. The training, validation, and test sets used in this work are mutually disjoint subsets of the overall dataset. The test set is completely independent and does not participate in model fitting or selection;

it is used exclusively to assess the model's generalization capability and serves as one of the most fundamental benchmarks for evaluating the MLP. In this work, 10% of the full dataset is reserved as the test set, and the remaining 90% forms the database pool, from which the training and validation sets are further defined. The validation set is used during fitting to monitor model performance and assist in model selection, helping to avoid overfitting. The final model is selected based on the minimal error of the validation set. In this work, 5% of the database pool is fixed as the validation set, and the remainder is used as the training set.

**Structure optimization and NEB calculations**

All structural optimizations were performed using ASE interfaced with CHGNet, VASP, large-scale atomic/molecular massively parallel simulator (LAMMPS)[70], and the corresponding potential backends (NNAP, NEP, DeePMD, and ReaxFF). DFT calculations were conducted at the PBE+D3(BJ)[71] level of theory. Geometry relaxations were carried out using the CG algorithm[72] in VASP and LAMMPS, while CHGNet employed the LBFGS optimizer[73] combined with UnitCellFilter[74] in ASE to relax both atomic positions and cell vectors. Convergence was achieved when the maximum atomic force was below 0.01 $eV·Å^{-1}$.

To obtain the pressure–volume relation, the DAAF crystal was optimized under hydrostatic pressures ranging from 0 to 10 GPa. The NEP89 was further fine-tuned using approximately 5000 DAAF structures sampled via HAL+FPS strategy within this pressure range, improving its accuracy in describing crystalline compression behavior.

All NEB calculations were performed in ASE using the climbing-image NEB[75] method with the FIRE optimizer[72]. Ten intermediate images were generated by the image dependent pair potential[76] interpolation scheme, ensuring smooth atomic displacements between fully optimized initial and final states. The NEB runs were considered converged when the maximum force on all images dropped below 0.05 $eV·Å^{-1}$.

**Assessment of MD stability**

To assess convergence during the crystal stage, we evaluated the MD stability of NNAP models trained at successive iterations. For each iteration, 50 biased MD simulations were performed in the NVT ensemble with different random seeds, each up to $10^4$ steps. Stability was judged from the first peak of the radial distribution function. A trajectory was classified as stable if both (i) the peak position remained within a tolerance window of 0.5 Å around its reference value from AIMD, and (ii) the normalized peak height deviated by less than 20% throughout the run. The average number of stable steps over the 50 runs was used as the robustness metric.

If enlarging the dataset no longer improved this metric—i.e., NNAP models trained with additional configurations exhibited converged stability behavior under multiple bias strengths—the crystal stage was deemed converged and the workflow proceeded to the next stage.

**Species identification**

Species identification was performed from a bonding network constructed using covalent-radii-based distance cutoffs by frame along NNAP-driven MD trajectories. Two atoms were considered bonded if their separation was smaller than a pair-specific cutoff, defined as the sum of their covalent radii with an empirical scaling factor. Customized distance thresholds were applied for hydrogen-related bonds (H–N, H–O, H–C, and H–H) to prevent spurious assignment of bond breaking or formation under high temperature conditions. The cutoffs were tuned by monitoring bond-length distributions from reference NNAP trajectories at 300 K, yielding thresholds of 1.38

Å for H–N, 1.24 Å for H–O, 1.30 Å for H–C, and 0.85 Å for H–H. A molecule was defined as a bond-connected set of atoms, and molecule with the same elemental composition were initially assigned to the same species.

To distinguish DAAF-specific reaction pathways, additional topology-based criteria were introduced for selected compositions. For molecule with the composition $C_4N_8O_3H_4$, the number of five-membered rings was evaluated, and structures with fewer than two five-membered rings were identified as RO. The ring-opening position was further determined by comparing the instantaneous bond network of RO with the reference topology of intact DAAF. Specifically, all bonds belonging to the two furazan rings in intact DAAF were predefined, and the missing bond in the RO was assigned as the scission site. When multiple original furazan-ring bonds were absent, or when one furazan unit had evolved beyond simple single-bond ring opening into an open-chain or rearranged N/O-rich fragment while the other furazan ring remained intact, the molecule was no longer classified as a RO species. Instead, such structures were collectively classified as FEP. For specie with the composition $C_4N_8O_3H_5$, the local bonding environment of the extra H atom was further analyzed to determine its attachment site, thereby distinguishing different HTr configurations. Other species only appearing more than 20 times and persisting for at least ten consecutive detection frames were retained for subsequent statistical analyses.

**Pyrolysis simulations**

Pyrolysis simulations were performed using NNAP on a supercell of DAAF (640 molecules) under the NVT ensemble at 1,750–2,750 K in increments of 250 K, each for 50 ps with a timestep of 0.1 fs. Complementary AIMD simulations were carried out in VASP at 3000 K for 15 ps on a smaller orthogonalized supercell containing eight DAAF molecules, using the same timestep.

The activation energy ($E_a$) for DAAF pyrolysis was extracted from temperature-dependent reaction rates $k(T)$, which was determined by fitting the decay of intact DAAF molecules to a first-order exponential model, and $E_a$ was subsequently obtained from the Arrhenius relation:

$$k\left(T\right) = A\exp\left(-\frac{E_{\mathrm{a}}}{RT}\right) \tag{8}$$

where $A$ is the pre-exponential factor, $R$ is the gas constant.

**Shock loading and Hugoniot curve**

To generate representative initial configurations for HAL sampling in the shock stage, DFT-level MSST simulations were performed in CP2K[77] on an orthogonalized supercell containing eight DAAF molecules. The simulations employed the PBE+D3(BJ), with a plane-wave cutoff of 500 Ry and a relative cutoff of 50 Ry. Shocks were applied along the $x$, $y$, and $z$ crystallographic axes, with a relatively high shock velocity of 10 km·s$^{-1}$ to efficiently capture the early-stage structural and chemical evolution. From these short-time DFT-level MSST trajectories, representative configurations corresponding to the onset of compression, maximum compression, and incipient chemical reactions were extracted. These configurations served as initial states for HAL sampling in the shock stage, ensuring that data generation remained anchored to physically meaningful configurations.

NNAP shock simulations were performed using LAMMPS on an orthogonalized DAAF supercell containing 640 molecules, subjected to loading along the $x$, $y$ and $z$ directions. The MSST parameters were set to $q$ = 400 and $mu$ = 1000, expressed in the LAMMPS metal unit system. By varying the $U_s$ and recording the corresponding $U_p$ after the system reached steady state, a series of $U_p$-$U_s$ data points were obtained and fitted to construct the Hugoniot curve.

**Conflicts of interest**
There are no conflicts to declare.

**Data availability**
The corresponding datasets will be made available upon acceptance of this work. Additional data that support the findings of this study are available from the corresponding author upon reasonable request.

**Author Contribution**
R.S. and P.F.G. conceptualized and designed the research. G.C.D., Z.Y.T., Q.A.L., L.H.R., R.S., and P.F.G. conducted the preliminary investigation. G.C.D., L.H.R., L.W., Y.X.C., R.S., and P.F.G. analyzed the data and interpreted the results. G.C.D., L.W., R.S., and P.F.G. jointly wrote the manuscript. L.W., B.X., J.L., S.Z., X.T.L., L.H.T., G.C.Y., J.K., R.S., and P.F.G all contributed to reading and editing the manuscript and approved the manuscript.

**Acknowledgements**
This work was supported by the Advanced Materials-National Science and Technology Major Project (Grant No. 2024ZD0606900), the Science Challenge Project. No.TZ2025013, the National Natural Science Foundation of China (Grants No. T2325004), and the Talent Hub for "AI+ New Materials" Basic Research.

# Supplementary Information

## Knowledge-Guided Construction of Machine Learning Potentials for Reactive Behavior in Energetic Materials

*Guanchen Dong[1,2,3,4], Huanrong Liu[1], Lin Wang[1], Yuxuan Chen[1,3], Zhengyu Tu[1,3], Qing-an Li[1,3], Bin Xu[1], Jian Li[1], Shan Zhang[1], Xiaotong Li[1], Linghua Tan[2], Guangcheng Yang[2,4], Jun Kang[3], Rui Su[5*], Pengfei Guan[1,3*]*

1. Advanced Interdisciplinary Science Research (AiR) Center, Ningbo Institute of Materials Technology and Engineering, Chinese Academy of Sciences, Ningbo 315201, China
2. School of Chemical Engineering, Nanjing University of Science and Technology, Nanjing, 210094, China.
3. Beijing Computational Science Research Center, Beijing 100193, China
4. China Academy of Engineering Physics·Institute of Chemical Materials, Mianyang, 621900, China.
5. Institute of Advanced Magnetic Materials, College of Materials & Environmental Engineering, Hangzhou Dianzi University, Hangzhou 310018, China

**Corresponding Author**

* Rui Su (surui@hdu.edu.cn); Pengfei Guan (pguan@nimte.ac.cn).

**Review on Initial Reaction of Representative and Furazan-Based Energetic Materials**

We compile a summary (Table S1) of reported initial reaction pathways for various energetic materials. Even for structurally similar furazan-based energetic materials, the primary initial reaction pathways are inconsistent. Faced with such complex nonlinear problems, it is difficult to predict the initial reaction pathways of different materials under various stimuli.

**Table S1.** Reported initial reaction pathways of representative and furazan-based energetic materials.

| Type | Initial reaction | Ref. |
|---|---|---|
| 2,4,6-Trinitrotoluene (**TNT**) | **C-$NO_2$ cleavage**; **nitro–nitrite rearrangement / H transfer** are alternative channels | [1] |
| 2,4,6-Triamino-1,3,5-trinitrobenzene (**TATB**) | **H transfer / dehydration** is commonly proposed; **C-$NO_2$ cleavage** and ring-related fragmentation are also discussed | [2–4] |
| 1,3,5-Trinitro-1,3,5-triazinane (**RDX**) | **N-$NO_2$ cleavage** | [5] |
| 1,3,5,7-Tetranitro-1,3,5,7-tetrazocane (**HMX**) | **N-$NO_2$ cleavage** and **C-H cleavage** | [6,7] |
| Hexanitrohexaazaisowurtzitane (**CL-20**) | **N-$NO_2$ cleavage** | [8] |
| Pentaerythritol tetranitrate (**PETN**) | **O-$NO_2$ homolysis** is favored; **HONO elimination** is competing but weaker | [9] |
| 1,1-Diamino-2,2-dinitroethylene (**FOX-7**) | **C-$NO_2$ cleavage** and **H transfer** are competing channels | [10,11] |
| 2,6-Diamino-3,5-dinitropyrazine-1-oxide (**LLM-105**) | **C-$NO_2$ cleavage** is the key initial channel; **H transfer** contributes at higher temperature | [12,13] |
| 3,4-Diaminofurazan (**DAF**) | **C(furazan ring)-N(furazan ring) cleavage** | [14] |
| 3,3′-Dinitro-4,4′-azoxyfurazan (**DNAF**) | **C(furazan ring)-N(azoxy) cleavage** | [15] |
| 3,4-bis(4-nitro-1,2,5-oxadiazol-3-yl)-1,2,5-oxadiazole (**BNFF-1**) | **Central heterocycle cleavage (RO)** is favored; **C-$NO_2$ cleavage** and **CONO** isomerization are competing channels | [16] |
| 3-(4-amino-1,2,5-oxadiazol-3-yl)-4-(4-nitro-1,2,5-oxadiazol-3-yl)-1,2,5-oxadiazole (**ANFF-1**) | **Central heterocycle cleavage (RO)** and **C-$NO_2$ cleavage** are competing channels | [16] |

i. The initial reactions listed in the table are conclusions drawn by different researchers using various approaches (thermal analysis experiments, DFT at different levels, or MD simulations), and these can be differed significantly under different types of stimuli. DAF, DNAF, BNFF-1, and ANFF-1 are furazan-based energetic materials.

**Machine learning potential (MLP) constructed using the AIMD+AL strategy**

AIMD-based dataset construction was carried out as follows. First, AIMD simulations were performed in both the NPT and NVT ensembles at 300 K for 10 ps on a series of 2×1×1 DAAF supercells optimized by density functional theory under pressures of 0 to 10 GPa with a step of 1 GPa. From each trajectory, 500 uniformly spaced configurations were selected to train the initial MLPs. Then, using an iterative procedure similar to that described in the main text, 5 ps AIMD simulations were performed in the NVT ensemble at temperatures from 100 to 3100 K with a temperature interval of 500 K, and configurations recorded every 0.01 ps. In each iteration, configurations were selected according to the uncertainty $U$ as in the main text, resulting in a final dataset of 7,806 configurations. The final MLP trained on this dataset using the same sphmulti basis set was referred to as "NNAP_AIMD".

**Knowledge-guided initial configuration selection and iterative criteria setting**

We propose framework for the knowledge-guided construction of machine learning potentials is shown in Fig. S1. Rather than relentlessly pursuing broad dataset, we decompose the application space of DAAF into the key physical and chemical events that matter in practice and build a target-driven dataset around these events. The goal is to steer optimization toward reproducing the dynamics of the key physical and chemical events of interest.

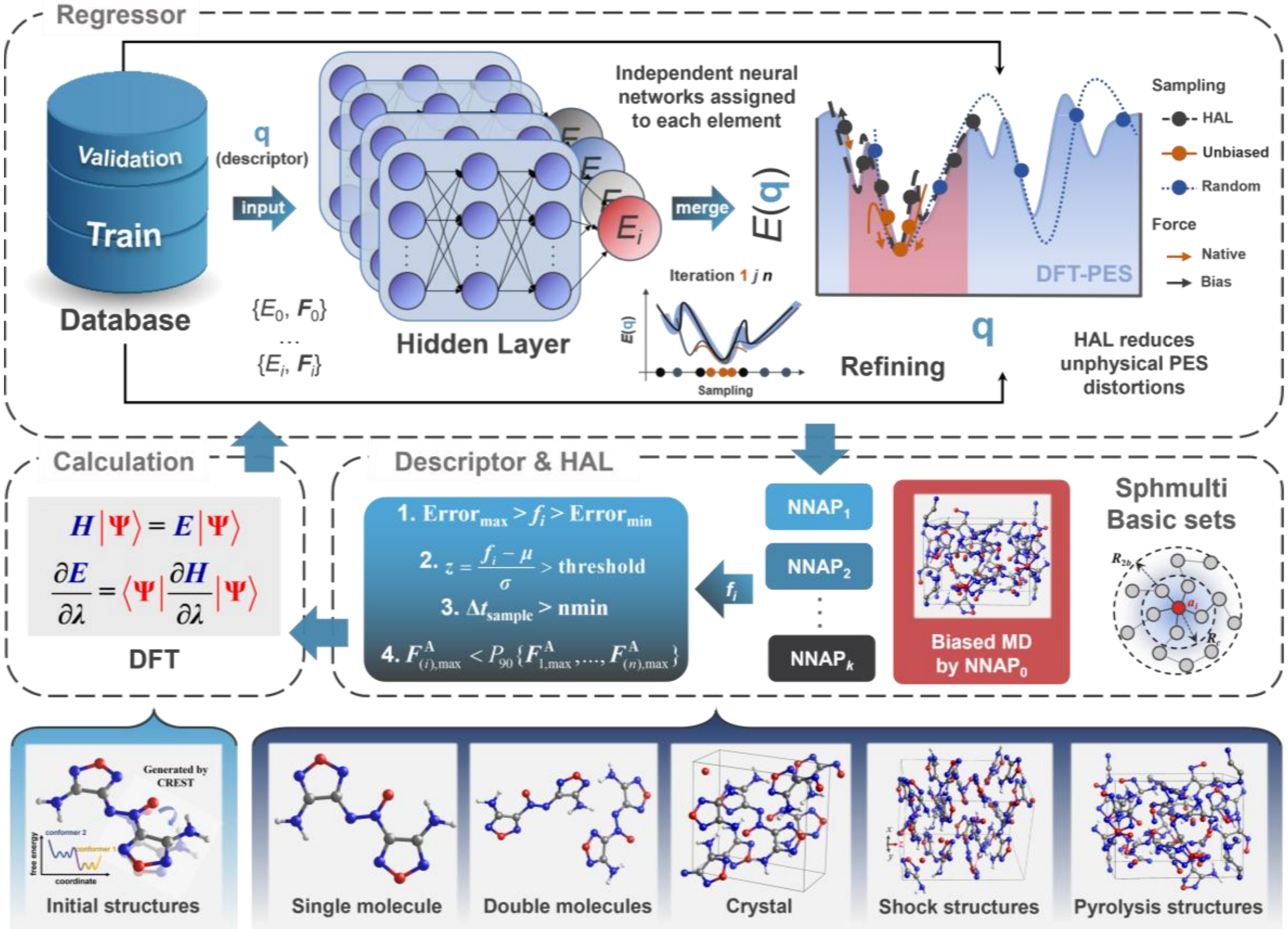

**Fig. S1.** Knowledge-guided machine learning potential framework schematic. Sampling and model refinement are organized into four stages—single molecule, crystal, pyrolysis, and shock—anchored to configurations representing key events. Each stage proceeds iteratively until its convergence criterion, defined by energy barriers, stability metrics, or experimental consistency, is satisfied. HAL and FPS ensure the efficiency of the overall sampling process.

The details of the four sampling stages for DAAF are given below.

(i) Molecule stage.

This stage addresses the most fundamental molecule configurational space. Since external loading in molecular crystals may induce not only packing rearrangements but also intramolecular deformations, reliable description at the molecular level must first be established. This stage therefore performs HAL sampling around multiple local minima, which were identified through CREST[17], performed searches of the isolated DAAF molecule using the GFN2-xtb[18] method. Based on chemical knowledge, one of the most characteristic molecular features of DAAF is the cis–trans isomerization process[19]. Iterative sufficiency criteria are reached when the error between NNAP and DFT in the isomerization barrier becomes smaller than MAE_E on the test set.

(ii) Crystal stage.

External loading can induce diverse changes in molecular packing, and the sampling therefore needs to extend from isolated molecules to collective structural responses in the crystal environment. Specifically, biased MD sampling is carried out from the DAAF crystal structure at different temperatures and pressures. Initial DAAF unit cell was experimentally determined under ambient conditions[20]. For this stage, the initial sampling configurations were taken from a series of 2×1×1 supercells optimized by DFT at different pressures. Since the objective of this stage is to establish the stability of crystal-level dynamical behavior, sampling sufficiency is assessed using an MD stability criterion. Although the specific rare events may vary, the overall dynamical behavior should exhibit statistical convergence.

(iii) Pyrolysis stage.

The continuous evolution of the local environment during pyrolysis gives rise to diverse initial reaction pathways. To capture the dominant reaction pathways in pyrolysis, AIMD simulations are first performed on a 2×2×1 DAAF supercell at 3000 K in the NVT ensemble, and configurations exhibiting significantly slowed electronic-step convergence are selected as representative structures near key bond-breaking events. HAL sampling is then carried out around these configurations. Using the NNAPs generated at each iteration, large-scale pyrolysis MD simulations are further performed at different temperatures, and the $E_a$ is determined from the decomposition rates of DAAF. This stage is regarded as sufficiently sampled when the resulting $E_a$ falls within the experimentally reported range[21,22].

(iv) Shock stage.

Similar to pyrolysis, the shock stage addresses a complex reactive process, but here the dynamics are jointly driven by thermal excitation and mechanical deformation, and is therefore placed at the final stage of the progressive framework. Specifically, AIMD-MSST simulations are carried out, and configurations near key events are selected according to significant changes in electronic-step convergence time and system volume, followed by HAL sampling. Sampling at this stage is considered sufficient when the Hugoniot curve predicted by the resulting NNAP agrees with the experimental Hugoniot[23].

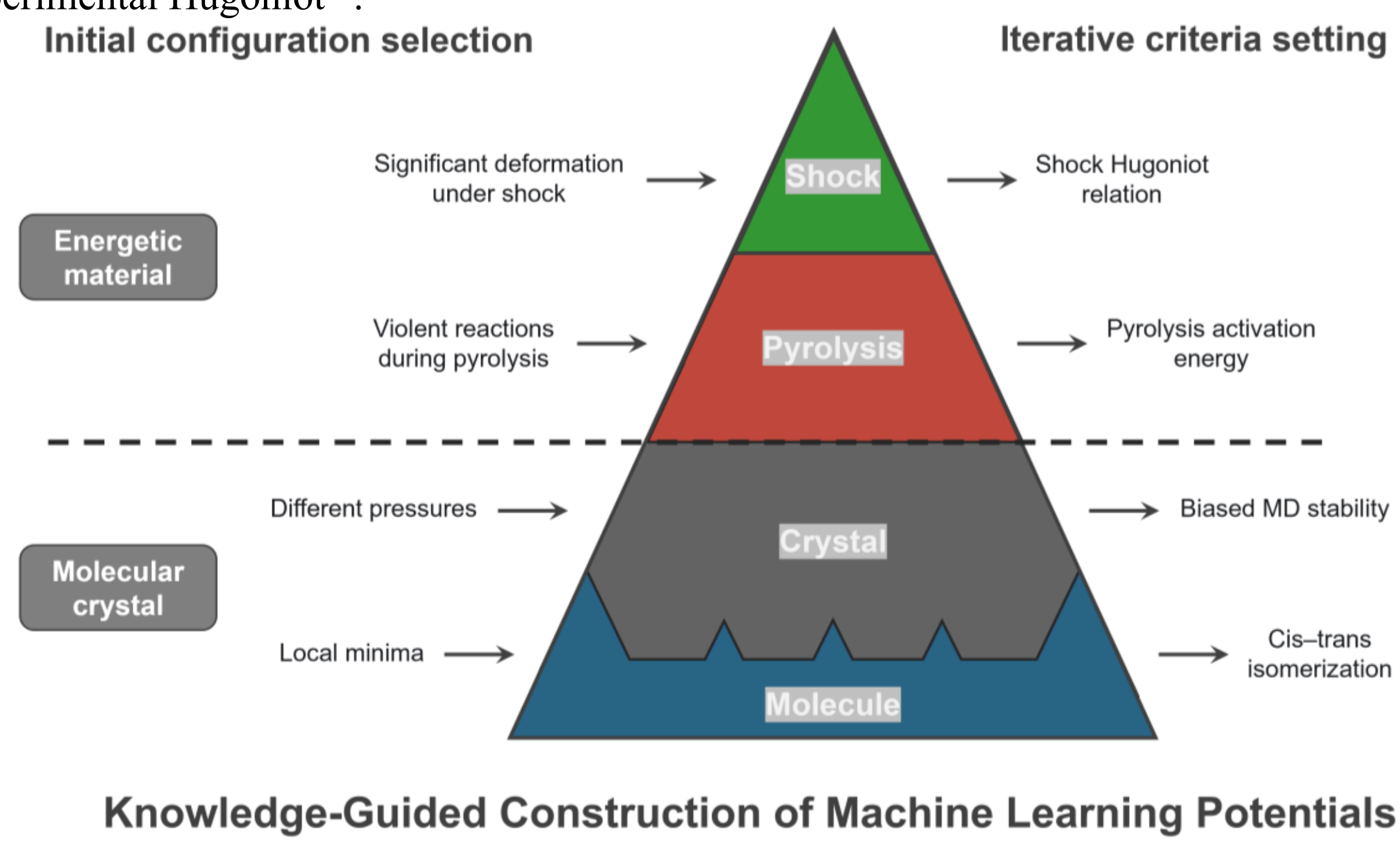

**Fig. S2.** The NNAP is constructed using a progressive knowledge-guided framework. First, key events are defined based on knowledge of the molecule and crystal stage, and the relevant conformations are selected. At the same time, the criteria for determining when the iteration for that stage is complete are established based on the key properties corresponding to those key events. The process then moves on to the pyrolysis and shock stages associated with energetic materials. Once all the criteria have been met, the final NNAP is obtained.

MLPs that exhibit low errors may still fail to reproduce the rational physical events. The energy/force errors of the different potentials are shown in Fig. 3a-d. Notably, the NNAP_AIMD exhibits a 57.2% lower force errors of mean absolute error (MAE_F) than the NNAP, although force errors are more directly relevant to dynamical simulations. However, the radial distribution function (RDF), results in Fig. 3e, f and Table S2, indicate that the NNAP_AIMD does not adequately reproduce the dynamic structural information of the DAAF molecular crystal at 300 K. Moreover, the conventional AIMD+AL sampling spreads uniformly in the principal component space of descriptors, which poorly covers the critical regions associated with shock and pyrolysis (Fig. S3).

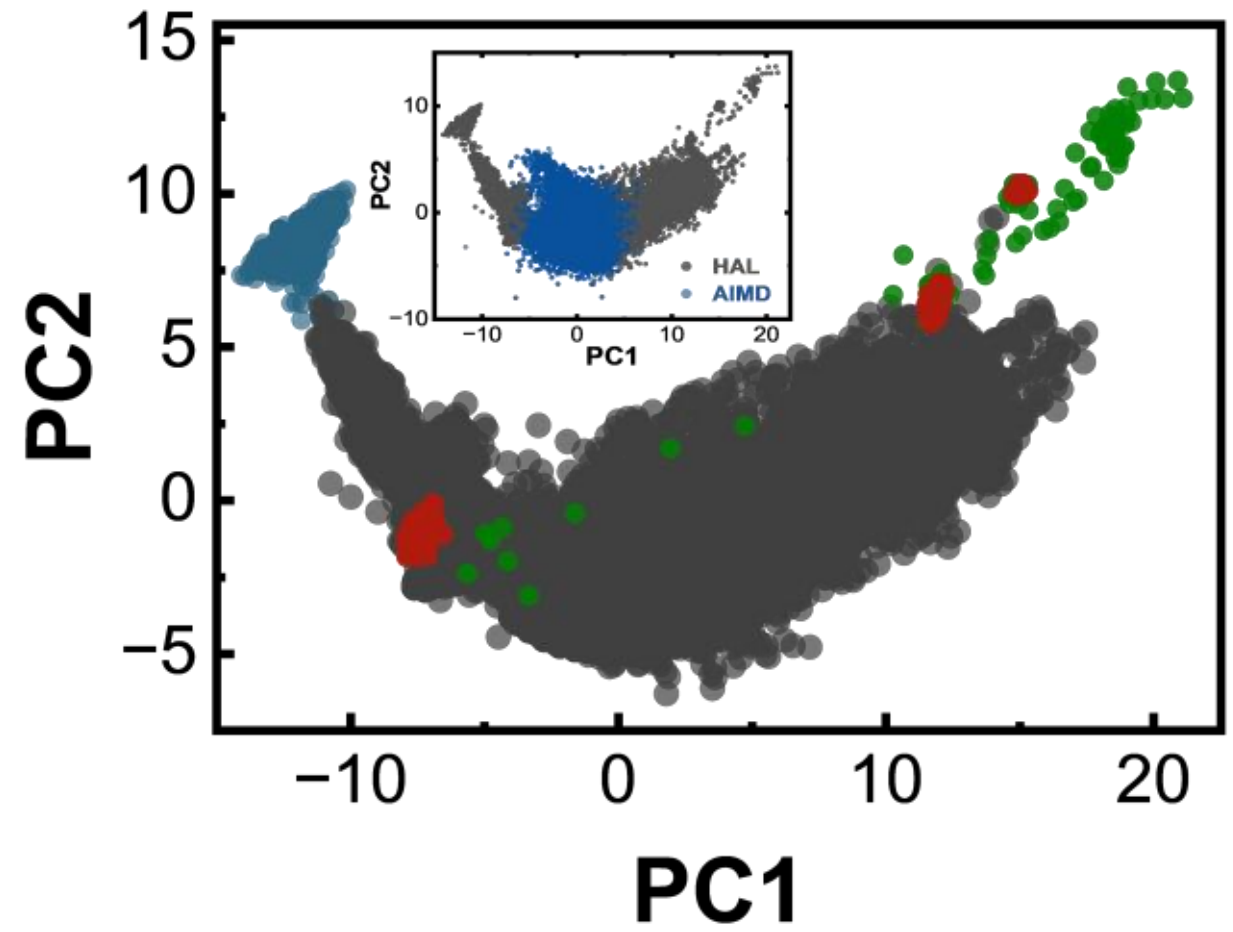

**Fig. S3.** Descriptor-space principal component analysis (PCA) of the dataset sampled by our framework (colors denote provenance: navy—single-molecule, gray—crystal, red—pyrolysis-related, green—shock-related). Inset: conventional AIMD+AL sampling (blue points) does not cover neighborhoods associated with related key events.

Specifically, we compared the performance of the NNAP developed in this work, a potential trained using the AIMD+AL sampling strategy (NNAP_AIMD), an NNAP trained with the sph basis set ($NNAP_{sph}$), an NNAP trained with the sphl basis set ($NNAP_{sphl}$) and a ReaxFF parameterized for DAAF (ReaxFF)[24] in describing the structural and vibrational properties of DAAF (Fig. 3e-f). All calculations followed the same protocol: a 50 ps NPT simulation was first carried out at 300 K and 0 pressure to obtain the RDF, followed by an additional 10 ps NVE simulation from which the vibrational density of states (VDOS) was computed using atomic velocity information. ReaxFF fails to reproduce the structural information of DAAF for $r \leq 2.5$ Å, with a MAE of 0.458 in $g(r)$ relative to DFT (MAE_$g(r)$). The potential constructed from AIMD sampling shows some improvement, but still exhibits significant deviations for $2.5 \leq r \leq 8.0$ Å (MAE_$g(r)$ = 0.148), and a similar issue is observed for the potential trained with the SPH basis set. A similar trend is found for the VDOS: although these potentials already show relatively small errors on the test set (MAE_E $\leq$ 0.01 eV and MAE_F $\leq$ 0.1 eV·Å$^{-1}$), their predictions still deviate noticeably from those of the NNAP developed in this work. The detailed comparison is summarized in Table S2.

**Table S2.** RDFs comparison of different potentials.

| **MAE_$g(r)$** | **$NNAP_{sphm}$** | **NNAP_AIMD** | **$NNAP_{sphl}$** | **$NNAP_{sph}$** | **ReaxFF** |
|---|---|---|---|---|---|
| ***r* < 2.5 Å** | **0.111** | 0.346 | 0.162 | 0.171 | 0.458 |
| **2.5 < *r* < 8.0 Å** | **0.0586** | 0.148 | 0.0884 | 0.0877 | 0.112 |

**Robustness Benchmarks of NNAP**

We first assessed predictive accuracy on the test set (see Fig. S4 for the predictive accuracy across the entire dataset).

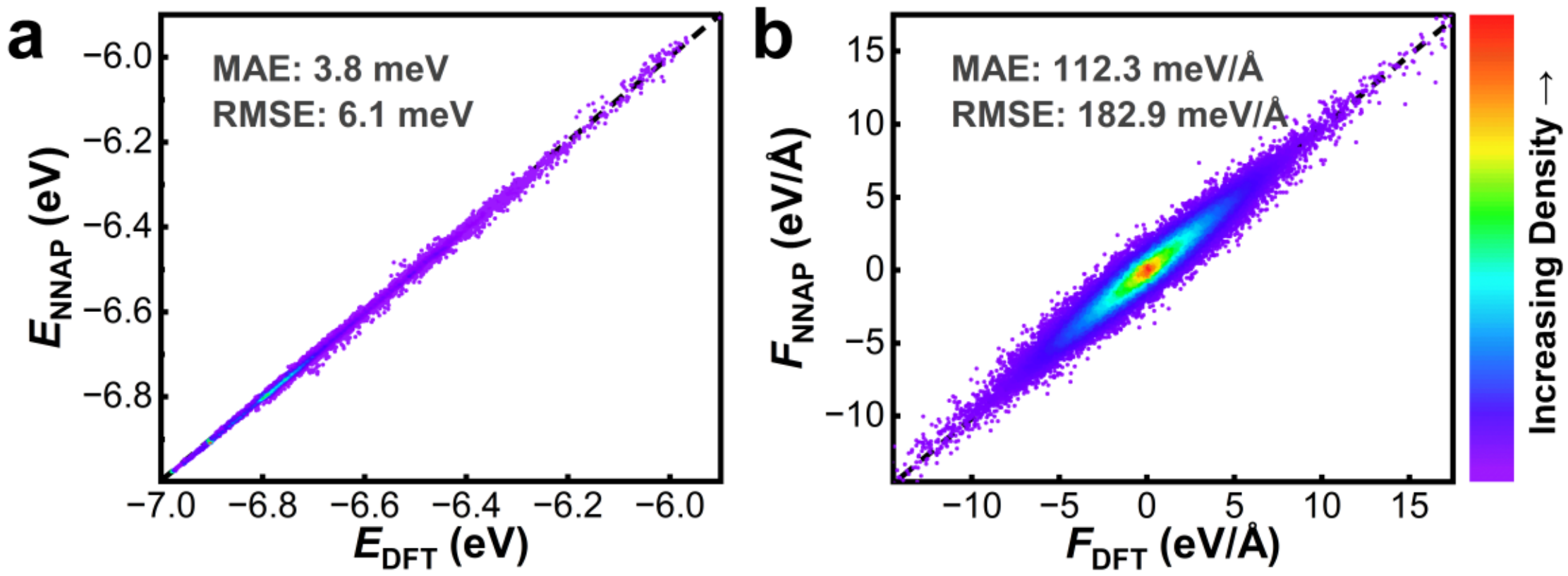

**Fig. S4.** Parity plots of (a) energies and (b) forces for the entire dataset configurations predicted by NNAP versus DFT, with corresponding MAE and RMSE values.

NNAP predictions attain tight agreement with DFT results, with energy errors of MAE = 3.7 meV and RMSE = 6.9 meV, and force errors of MAE = 112.5 meV·Å$^{-1}$ and RMSE = 189.5 meV·Å$^{-1}$ (Fig. S5a,b). Test set errors are comparable to the entire dataset errors, indicating no noticeable overfitting. The cell parameters are shown in Fig. S5c. NNAP reproduces experimental cell parameters[20] to within 1%, closely matching DFT. Moreover, the 0 K elastic constants obtained from NNAP agree with DFT (Fig. S5d), where $C_{ij}$ represents the stiffness constant for the linear coupling between strain components corresponding to indices *i* and *j*, with *i* and *j* corresponding to normal strains along the *a*, *b*, and *c* directions of DAAF unit cell, and shear strains like *ab*, *bc*, and *ca*. The relative error in $C_{25}$ arises from the difference between the DFT- and NNAP-predicted values, which are 3.0 and 5.9 GPa, respectively. The absolute errors of all the elastic constants predicted by NNAP compared to DFT calculations are all less than 4 GPa, confirming NNAP's ability to capture the short-range interactions and longer-range, and packing structure relevant mechanical responses.

The equation of state across multiple pressures is shown in Fig. S5e, and the predicted pressure-density relation tracks the experimental equation of state[25] across the entire range tested. In contrast, the DAAF-specific ReaxFF[24], EMFF2025 (trained on CHON energetic-material dataset based on DeePMD)[26], and fine-tuned NEP89[27] (details see Method) exhibit sizable deviations in these basic structural metrics of crystalline DAAF (Fig. S7b–f). To validate NNAP's ability to capture finite-temperature structure and mechanical properties, we first computed radial distribution functions from NNAP-MD at 300 K, which faithfully reproduce DFT results (Fig. S5f, and a more detailed comparison, including other potentials, is presented in Fig. 3e and Table S2.).

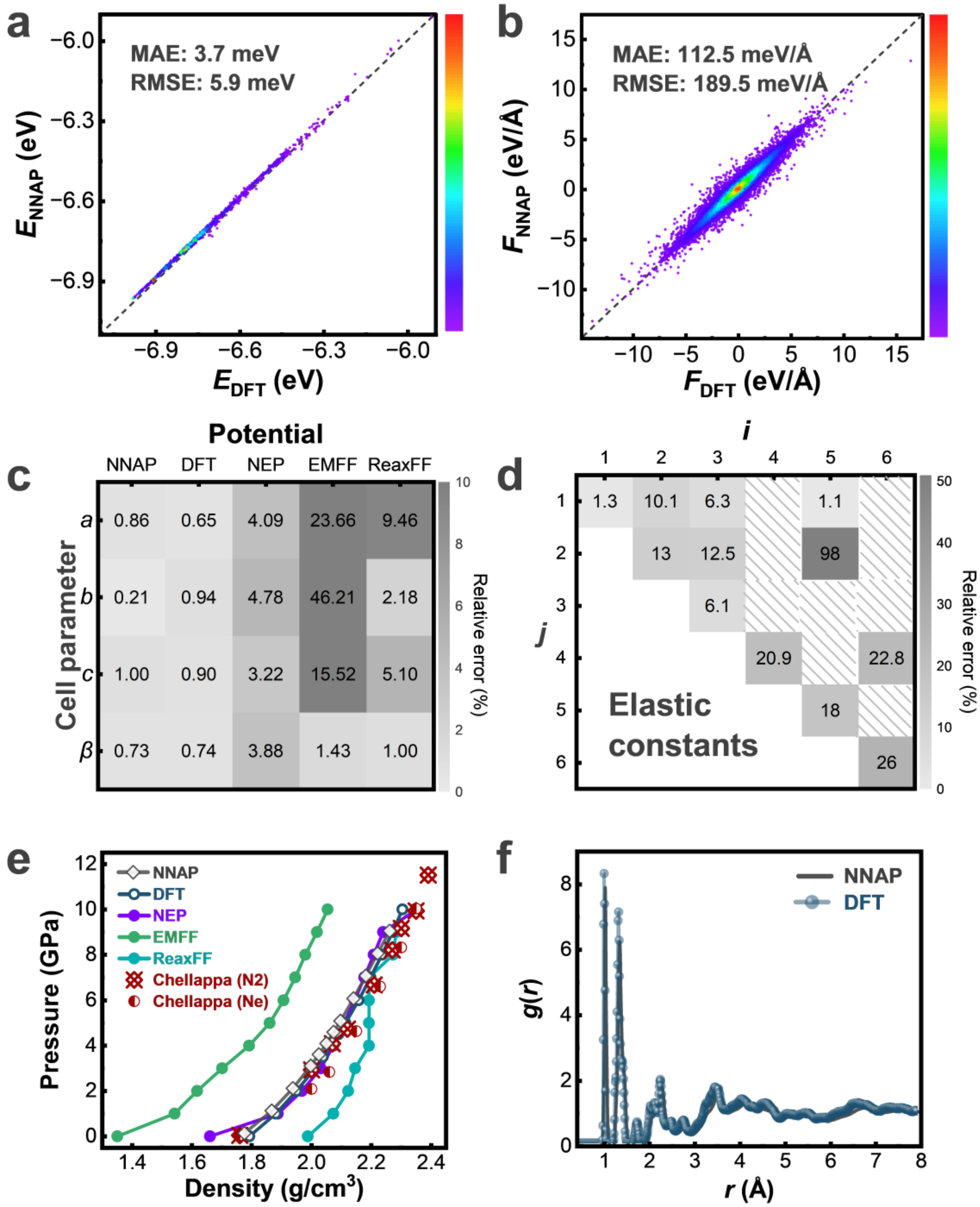

**Fig. S5.** Benchmarks of NNAP for crystalline DAAF. (a,b) Parity plots of energies and forces for test-set configurations predicted by NNAP versus DFT, with corresponding MAE and RMSE values. (c) Relative error of cell parameters[20] with respect to experimental values, predicted by NNAP, DFT, a fine-tuned NEP89 potential[27], a DeePMD trained on CHON energetic-material datasets[26], and a DAAF-specific ReaxFF[24]. (d) Relative error of elastic constants, predicted by NNAP, with respect to DFT values. The elastic constants of crystalline DAAF obtained from stress–strain fitting at 0 K using finite distortion method. (e) Equation of state: density–pressure relation of DAAF from NNAP, DFT, and other potentials, compared with experiment[25]. (f) Radial distribution function of a 2×2×2 DAAF supercell at 300 K from NNAP-MD and AIMD.

To further validate NNAP on dynamical properties, we examined its ability to predict vibrational features of crystalline DAAF at 300 K. As shown in Fig. S6, the vibrational density of states (VDOS) obtained from MD velocities closely reproduces the experimental IR and Raman peak positions[25], confirming that NNAP accurately captures the interatomic interactions of DAAF. Beyond the spectral information matching, a key advantage of a high-fidelity MLP is the ability to decompose and track complex spectrum using atomistic trajectories. By projecting the VDOS onto subspaces defined by internal coordinates (bond stretching, angle bending, et. al.), we assign overall vibrational features to specific motions with atomistic clarity.

For the C–N bonds connecting the furazan ring and the azoxy group (Fig. S6b), the Fourier transform of the projected autocorrelation function exhibits dominant peaks at 1451 $cm^{-1}$ and 1477 $cm^{-1}$, identifying them as the modes with the strongest C–N stretch participation. The dominant peak of C–N stretch nearer to the azoxy O is blue-shifted. The two distinct C–N stretches display strong coupling around 1261 $cm^{-1}$, indicative of symmetric and antisymmetric stretching combinations. Similarly, for the azoxy N=N[O] coordinates (Fig. S6b), the N=N stretch shows the highest mode intensity and is strongly coupled to the N–O stretch. Using the same projection method, we isolate the $–NH_2$ group signatures and clearly distinguish symmetric and antisymmetric stretches (Fig. S6c), the relative intensities of each projected vibrational modes consistent with VDOS (Fig. S6a).

These accurate vibrational modal projections capability provides a powerful tool for (i) atom-level interpretations for complex vibrational spectrum; (ii) allowing direct simulation and analysis of multiphonon up-pumping under stimuli loading.

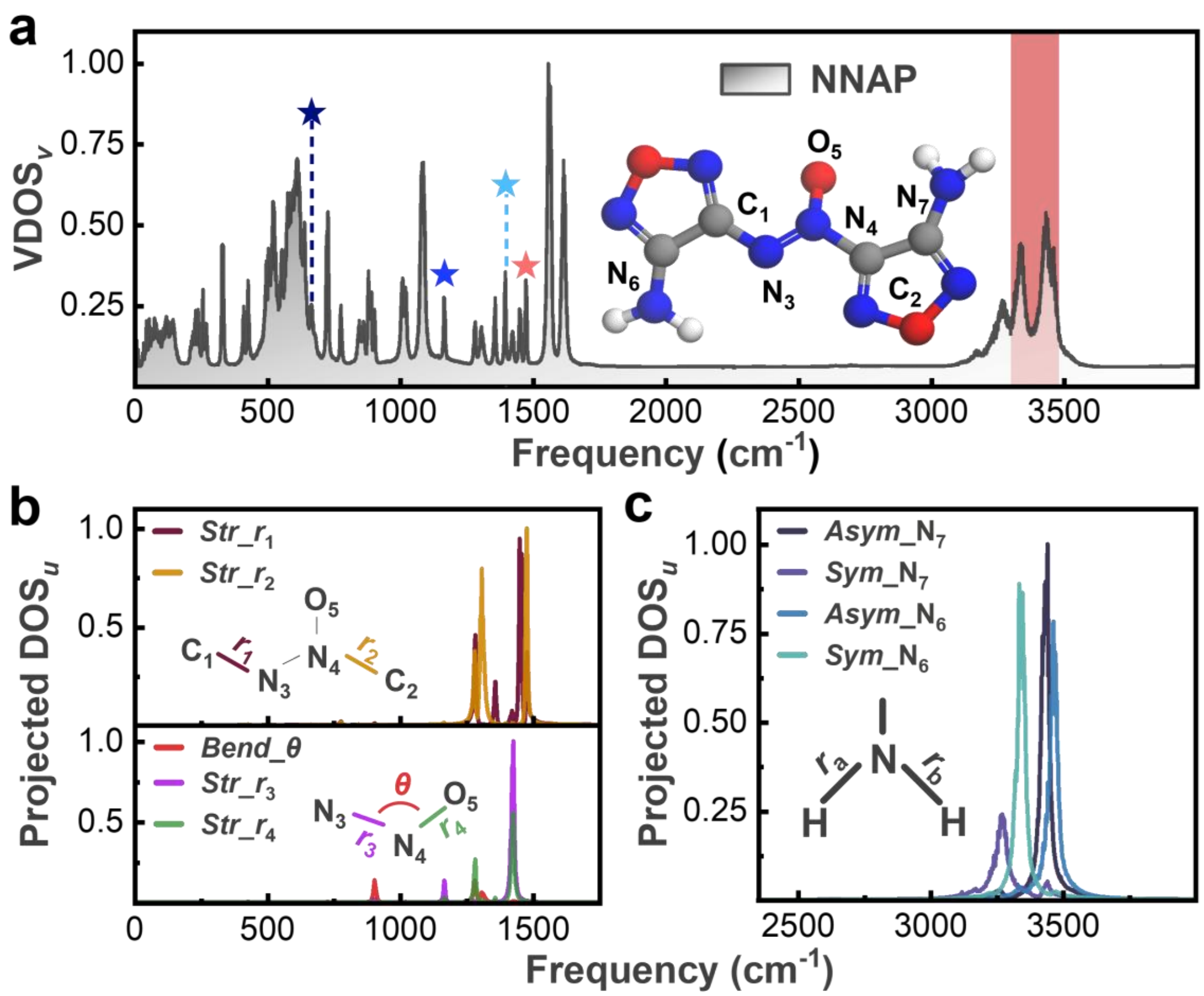

**Fig. S6.** (a) VDOS of crystalline DAAF at 300 K from NNAP-MD compared with experimental IR/Raman spectrum. Symbols denote vibrational modes from experimental report[25]: dark blue star—$\tau$(ON–H), blue star—$\rho(NH_2)$, light blue star—$\nu_s$(C=N), pink star—$\nu$(N=N), and red region highlighting vibrations associated with amino groups. (b) Internal-coordinate projections of stretching modes involving the C atoms of the furazan ring and the N atoms of the azoxy group, as well as bending/stretching modes of the azoxy group. (c) Internal-coordinate projections of symmetric and antisymmetric vibrations of the amino groups.

Finally, NNAP delivers the high accuracy in mechanical and structural behavior of DAAF without increasing NNAP's complexity significantly. In our throughput tests, NNAP is slightly faster than NEP89 (Fig. S7a). and with the knowledge-guided sampling, it achieves high fidelity while markedly reducing the cost of MLP training. In summary, these benchmarks demonstrate that NNAP not only captures short-range intermolecular interactions but also reliably reflects the longer-range properties that governs the macroscopic mechanical behavior of DAAF.

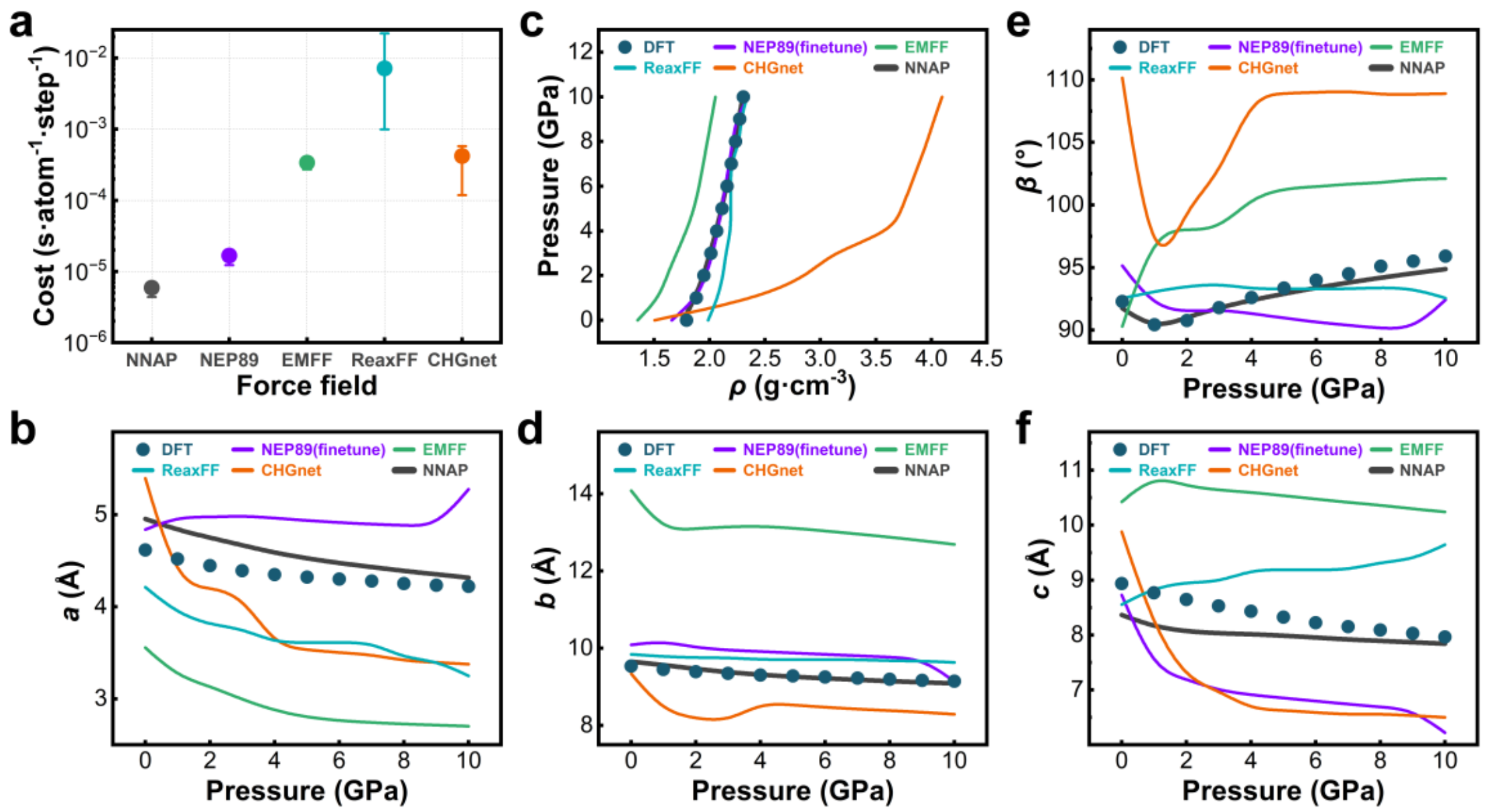

**Fig. S7.** (a) Computational cost comparison among different potential frameworks tested in LAMMPS. All simulations were performed on 56 CPU cores of an Intel® Xeon® Gold 6348 processor (2.60 GHz) under identical MD settings. (b-f) Equation of state: density–pressure relation and detail cell parameter of DAAF from NNAP, DFT, and other force fields.

**VDOS and mode projections**

VDOS and mode projections were obtained from MD trajectories of a DAAF supercell with 3,040 atoms (160 molecules). The system was equilibrated at 300 K and 0 GPa using 10 ps of NPT followed by 5 ps of NVT, and a subsequent 100 ps NVE production run was performed with positions and velocities saved every 4 fs.

The VDOS was computed based on the autocorrelation function (ACF):

$$C_v(t) = \frac{1}{N}\sum_{i=1}^{N}\frac{\left\langle \mathbf{v}_i(t_m)\cdot\mathbf{v}_i(t_m+t)\right\rangle_{\Delta T}}{\left\langle \mathbf{v}_i(t_m)\cdot\mathbf{v}_i(t_m)\right\rangle_{\Delta T}} \tag{9}$$

where averaging $\langle\cdot\rangle_{\Delta T}$ was performed over time windows of length $\Delta T$ = 20 ps with 50% overlap, defined as $t_m = m\cdot\Delta T/2$.

Then VDOS was obtained via Fourier transform:

$$S_v(\omega) = \frac{1}{M}\sum_{m=1}^{M}\int_{-\Delta T/2}^{\Delta T/2} C_v^{(m)}(t)\, e^{-i\omega t}\, dt \tag{10}$$

where $C_v^{(m)}(t)$ denotes the ACF within the $m$-th time window, and the normalized VDOS as

$$g(\omega) = \frac{S_v(\omega)}{\int_0^{\infty} S_v(\omega)\, d\omega} \tag{11}$$

The contribution of vibrational motions to the VDOS was evaluated by projecting the atomic velocities onto the corresponding internal coordinates. The symmetric and antisymmetric stretching mode of $NH_2$ group was defined by the natural internal coordinates[28].

$$u_{sym} = \frac{1}{\sqrt{2}}(r_a + r_b),\; u_{asym} = \frac{1}{\sqrt{2}}(r_a - r_b) \tag{12}$$

where $r_a$ and $r_b$ are the bond length of N–$H_1$ and N–$H_2$, respectively. Other internal coordinates, such as C–N stretching, was defined by the corresponding bond lengths or angles. After the projection, the ACF and projected DOS of each mode were calculated by the same procedure mentioned above. This internal coordinate projection offers the identification of VDOS from the contributions of specific vibrations.

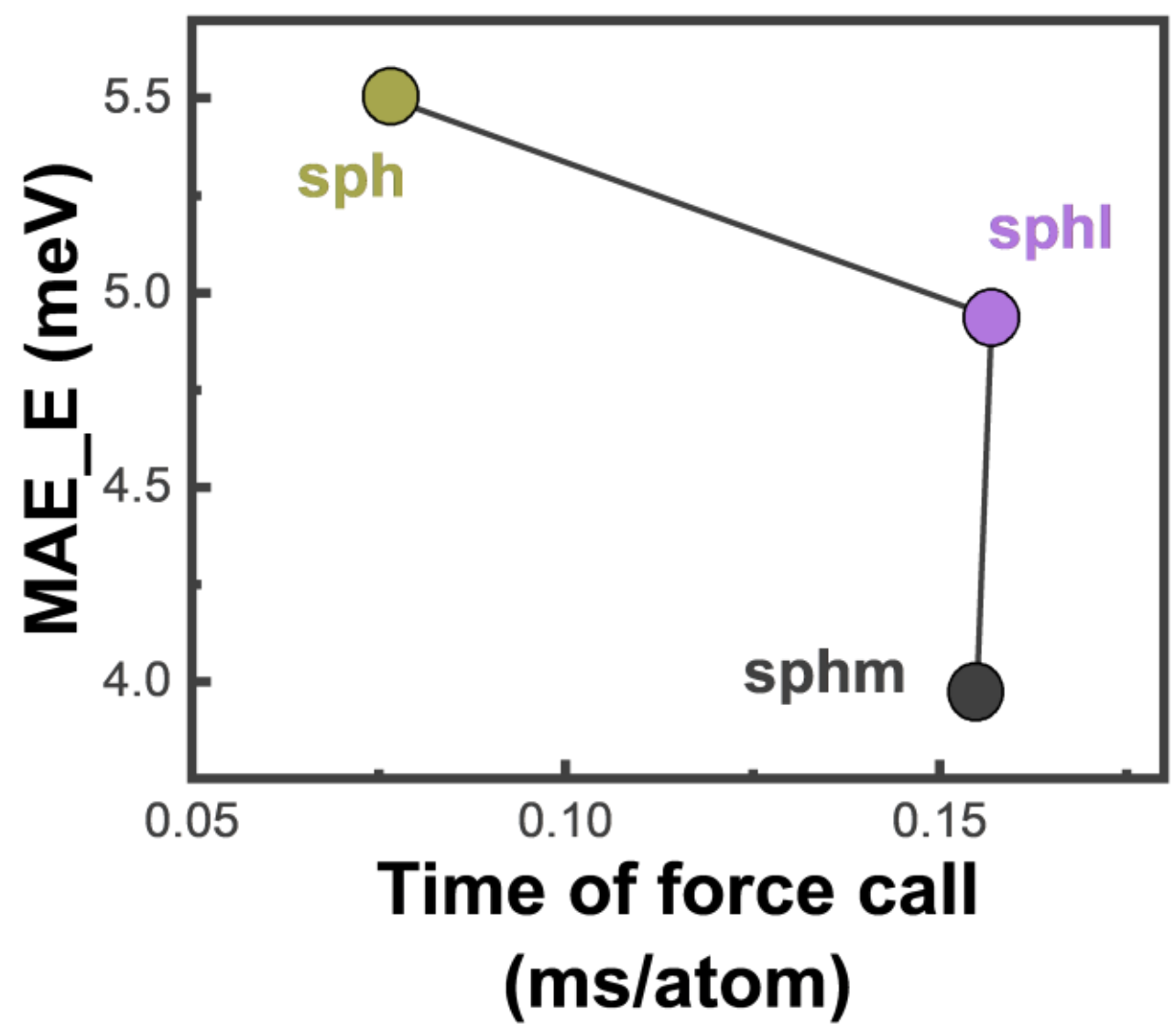

**Fig. S8.** Comparison of computational efficiency and accuracy for different basis sets of NNAP. The time of force call is measured using LAMMPS[29] on one CPU core of an Intel® Xeon® Gold 6348 processor at 2.60 GHz for a system containing 108 DAAF molecules. The MAE_E is obtained from 10-fold cross-validation on the test set for different basis sets. In general, a smaller time of force call corresponds to lower computational cost but reduced accuracy. Although the sph is computationally efficient, its accuracy is much lower than that of sphm. Increasing only the radial cutoff, as in sphl, increases the computational cost but improves the MAE_E by 10.1% compared with sph. More importantly, the improved representation of the weight terms in sphm is complementary to the increased radial channels, leading to a further 19.2% reduction in MAE_E compared with sphl, while maintaining a slightly lower time of force call. These results indicate that sphm achieves a better balance between computational efficiency and accuracy.

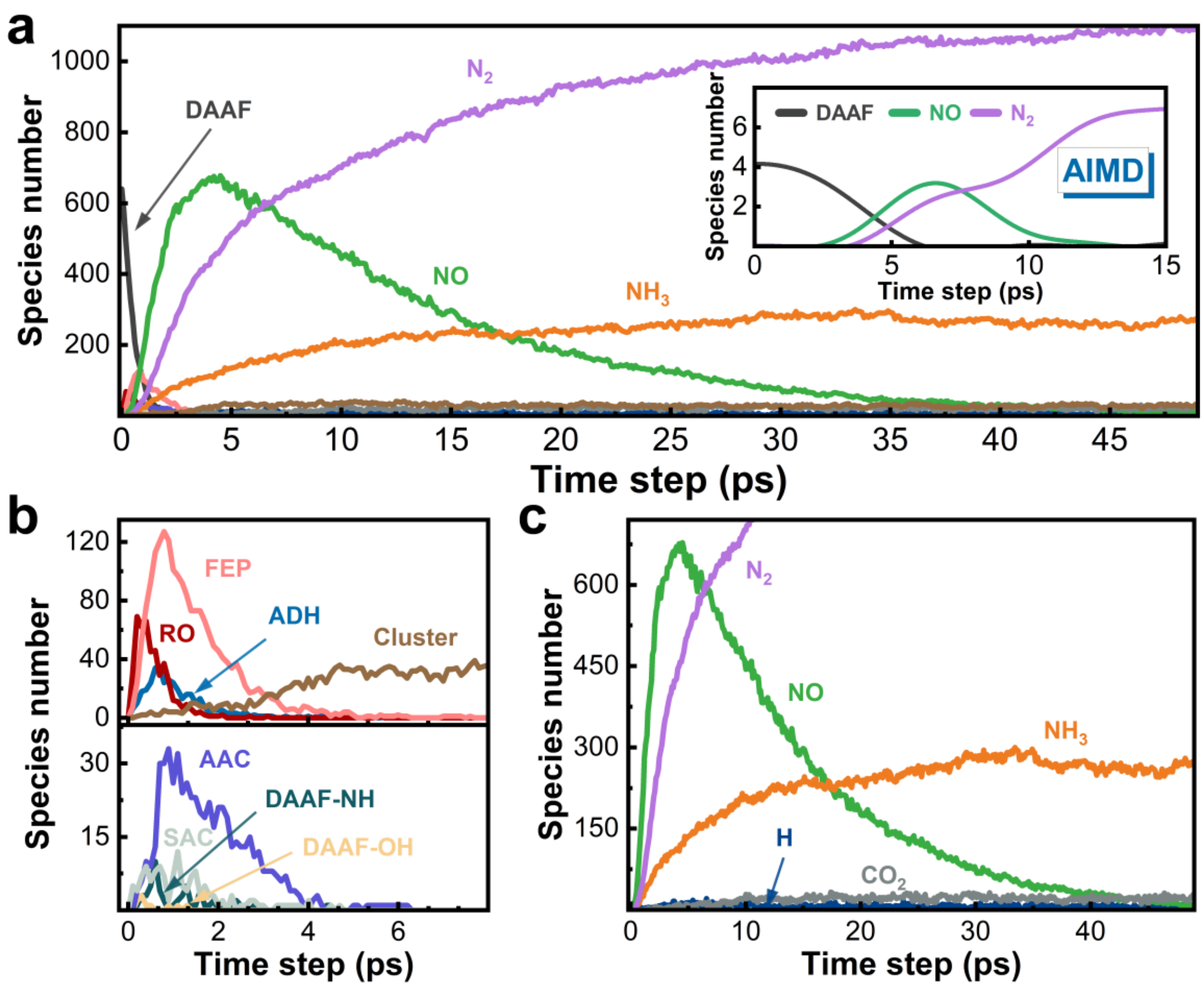

**Fig. S9.** Species evolution during pyrolysis of DAAF. (a) Time evolution of the major species during pyrolysis at 2500 K; inset compares with that of AIMD at 3000 K. (b) Early-time counts of key intermediates. The upper panel shows the evolution of initial events-related (including RO and HTr) key intermediates, while the lower panel tracks the subsequent species associated with further RO/HTr reactions. Furazan evolution product (FEP) denote a group of species that have evolved further following RO. Amino dehydrogenation (ADH) denotes DAAF after loss of one H atom from an amino group. DAAF-NH and DAAF-OH denote species with an additional H atom attached to a furazan-ring N atom, and the azoxy O atom, respectively. (c) Counts of the key small-molecule products during pyrolysis.

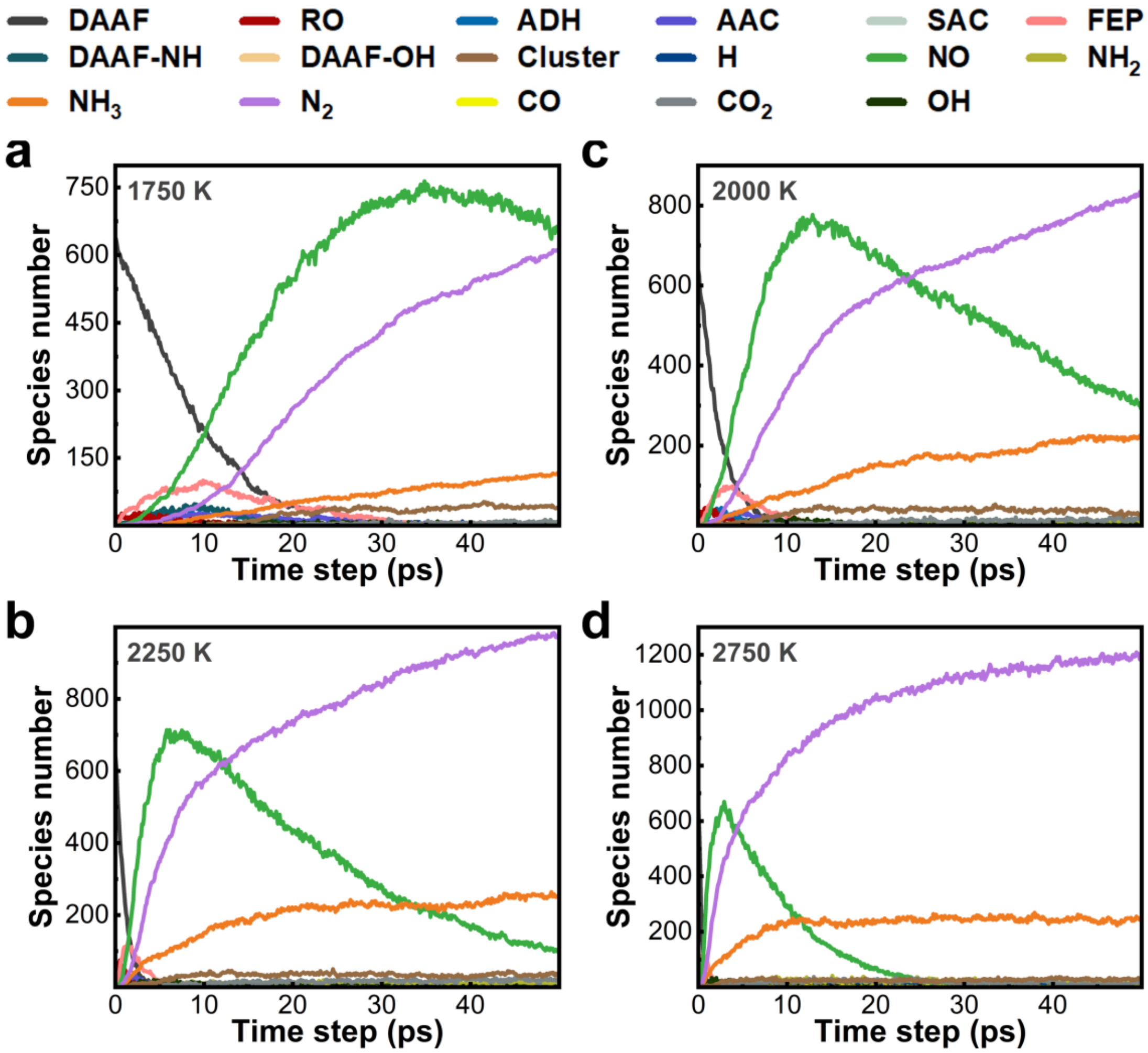

**Fig. S10.** Time evolution of the major species during pyrolysis at (a) 1750 K, (b) 2000 K, (c) 2250 K, and (d) 2750 K.

## Rationality test of NNAP in describing the shock process

To verify whether NNAP can accurately capture the overall evolution trend of DAAF under shock loading, we performed a 2 ps shock simulation in CP2K[30] (at the PBE-D3(BJ)/DZVP-MOLOPT-SR-GTH level, using a plane-wave cutoff of 500 Ry and a relative cutoff of 50 Ry) by applying a shock velocity $U_s$ of 12 km·s$^{-1}$ along the *x* direction, and extracted 50 configurations from the trajectory at equal time intervals. Energy calculations were then carried out for these sampled structures using both DFT (at the same computational level as that used for dataset labeling) and NNAP, and the variation of total energy with time was compared (Fig. S10). The results show that, although DFT exhibits more pronounced energy fluctuations, especially during highly reactive stages, the overall response trend predicted by NNAP remains highly consistent with that of DFT. Specifically, the per-atom MAE and RMSE over all sampled points are 0.152 and 0.156 eV, respectively. Further comparison of the energy gradients between adjacent frames shows that the per-atom MAE and RMSE of NNAP relative to DFT are only 3.77 and 32.3 meV, respectively, demonstrating that NNAP can reliably capture the overall reaction evolution trend under shock loading with DFT-level accuracy.

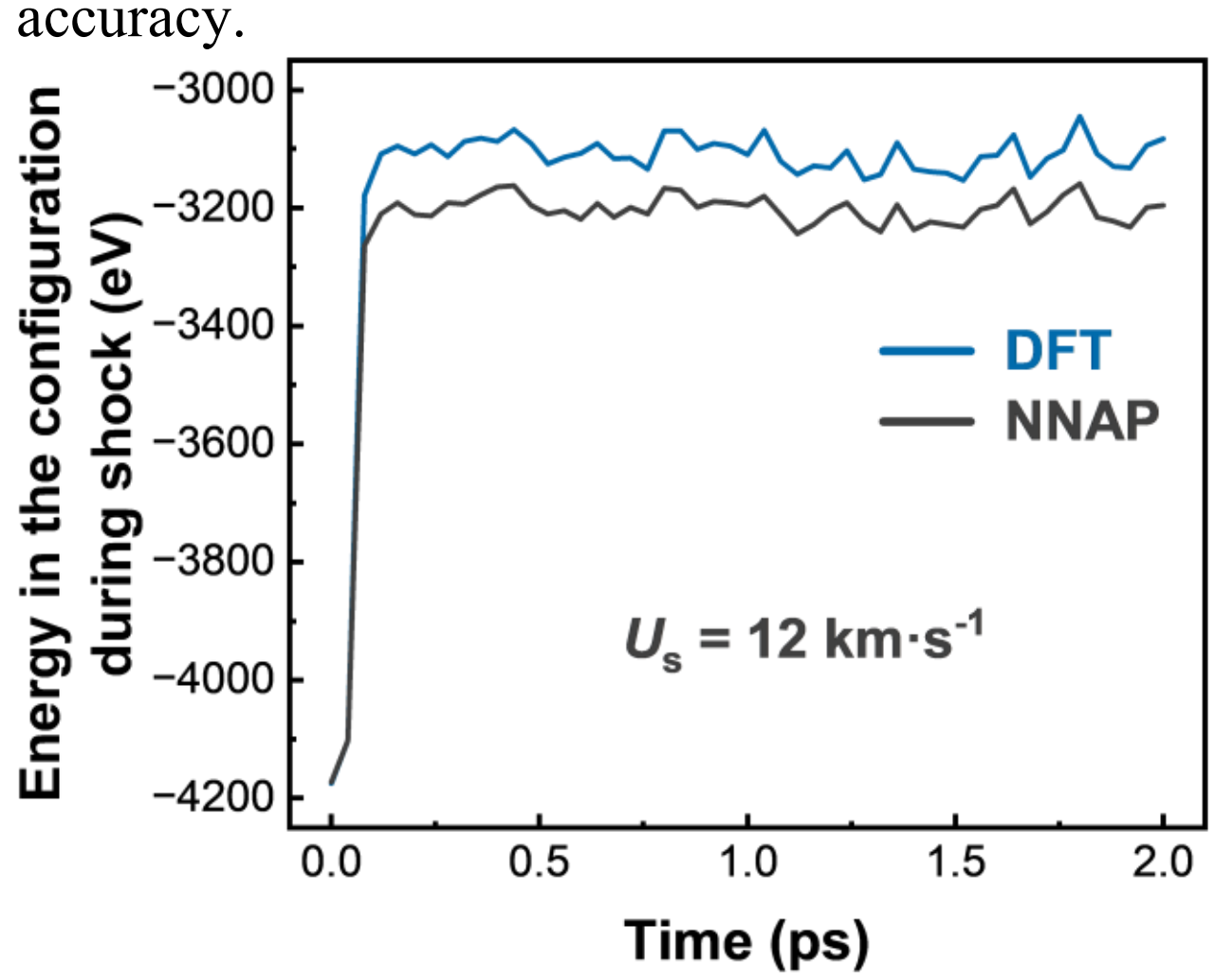

**Fig. S11.** Comparison of configuration energies during shock loading calculated by NNAP and DFT.

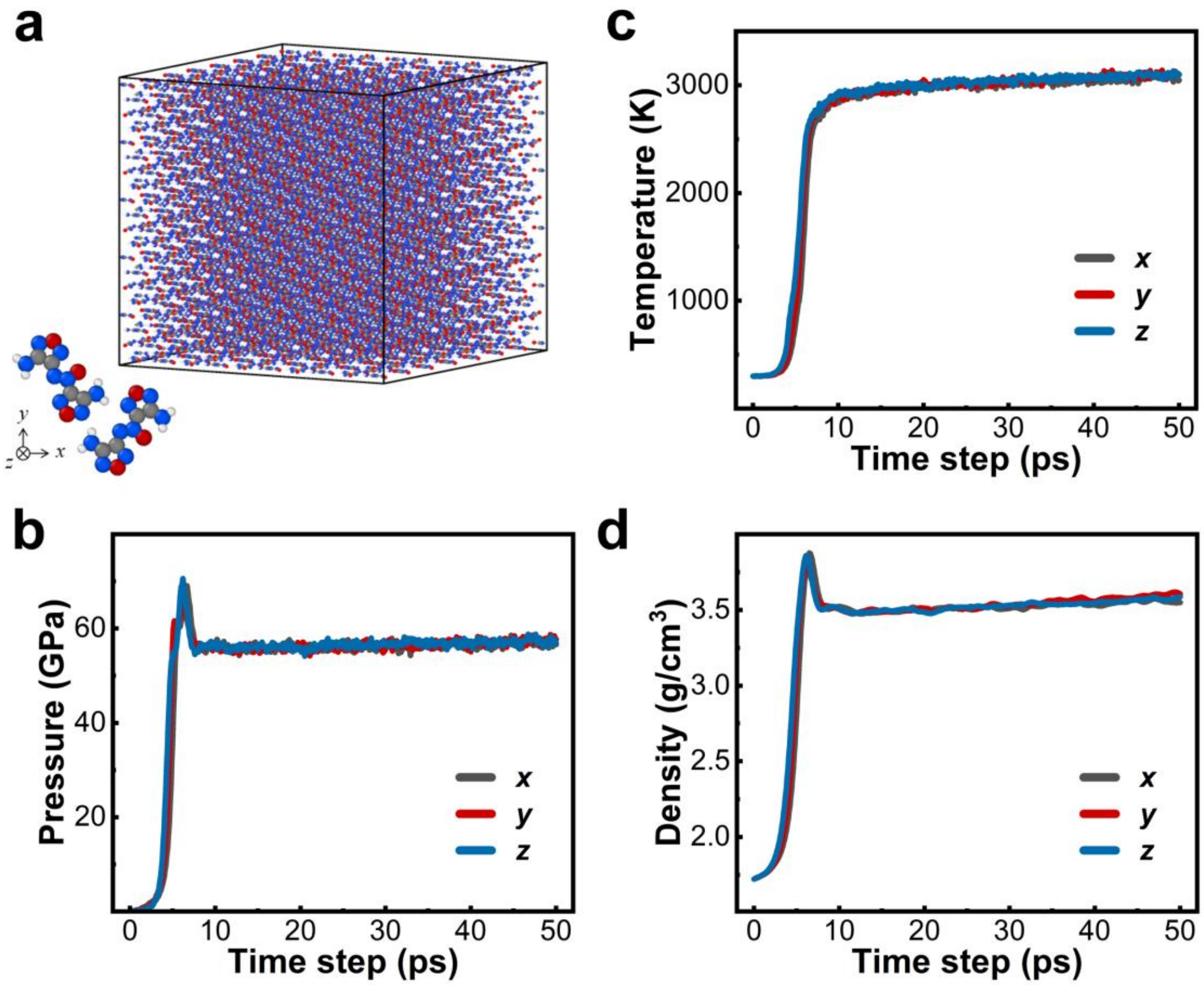

**Fig. S12.** (a) Atomistic representation of the orthogonalized DAAF supercell used for MSST shock simulations. (b–d) Time evolution of pressure, temperature, and density under shock loading along the x, y, and z crystallographic directions, respectively.

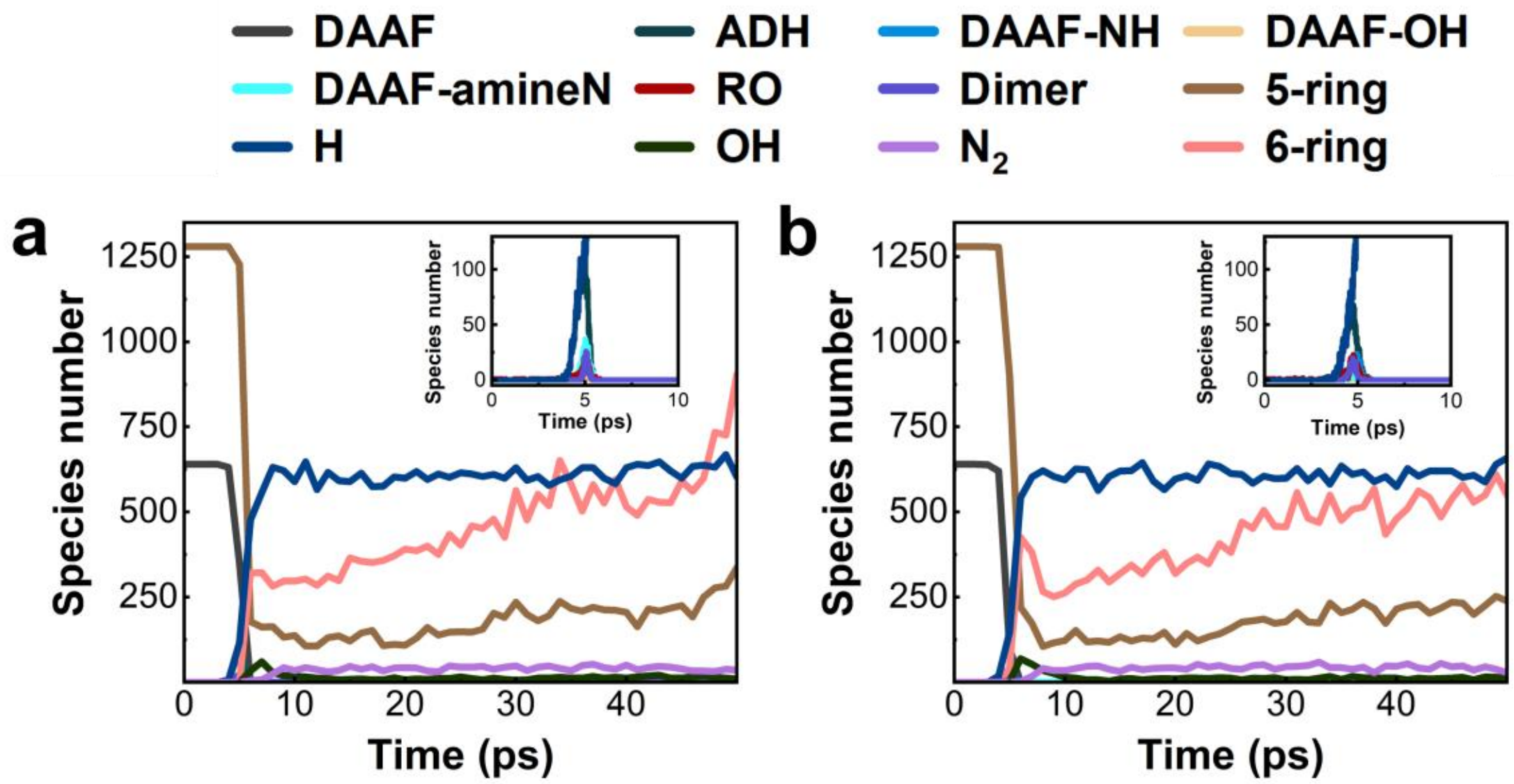

**Fig. S13.** Time evolution of the major species under shock loading along the (a) *y* and (b) *z* direction; inset shows detailed counts of key intermediate species.

**Table S3.** Quantitative comparison of NNAP performance in describing the four key events of DAAF.

| | $\boldsymbol{E_{iso}}$ (eV·atom$^{-1}$) | $\boldsymbol{E_a}$ (kJ·mol$^{-1}$) | $\boldsymbol{\lambda}$ | $\boldsymbol{c_0}$ (km·s$^{-1}$) |
|---|---|---|---|---|
| **NNAP** | 0.302 | 133 | 1.62 | 3.02 |
| **DFT** | 0.301 | - | - | - |
| **EXP** | - | 121-164 [21,22] | 1.79 [23] | 2.51 [23] |

i. where $\boldsymbol{E_{iso}}$ is the energy barrier for cis-trans isomerization; $\boldsymbol{E_a}$ is the activation energy for thermal decomposition; $\boldsymbol{c_0}$ is the bulk sound speed at zero pressure; and $\boldsymbol{\lambda}$ is the slope parameter in the linear Hugoniot relation $\boldsymbol{U_s = c_0 + \lambda U_p}$; $\boldsymbol{U_s}$ and $\boldsymbol{U_p}$ are the shock and particle velocities, respectively.